\documentclass[twocolumn, superscriptaddress, amsmath, amssymb, pre]{revtex4-1}

\usepackage{tikz}
\usetikzlibrary{decorations.pathmorphing}
\usetikzlibrary{decorations.markings}
\usetikzlibrary{arrows,shapes,snakes,automata,backgrounds,petri}
\usetikzlibrary{calc}
\usetikzlibrary{patterns}
\usetikzlibrary{decorations.text}
\tikzset{
xxtsubstrate/.style={decorate, 
line width=1pt,
draw=olive, 
decoration=snake, 
segment amplitude=0.75mm, 
line after snake=0.25mm,
line before snake=0.25mm
},
tsubstrate/.style={decorate, 
line width=1pt,
draw=olive, 
decoration=snake, 
segment amplitude=0.5mm, 
segment length=5pt,
segment amplitude=0.2mm, 
line after snake=1mm,
line before snake=1mm
},
Bsubstrate/.style={decorate, 
line width=1pt,
draw=olive, 
decoration=snake,
segment length=5pt,
segment aspect=0,
segment amplitude=0.5mm, 
line after snake=0mm,
line before snake=0mm
},
substrate/.style={decorate, 
line width=1pt,
draw=olive, 
decoration=snake, 
segment length=5pt,
segment amplitude=0.5mm, 
line after snake=0.5mm,
line before snake=0.5mm
},
activity/.style={very thick,draw=red,postaction={decorate},
decoration={markings,mark=at position .5 with
{\arrow[draw=red]{>}}}},
tactivity/.style={thick,draw=red,postaction={decorate},
decoration={markings,mark=at position .5 with
{\arrow[draw=red]{>}}}},
tEPSactivity/.style={thick,draw=red,postaction={decorate},
decoration={markings,mark=at position .55 with
{\arrow[draw=red]{>}}}},
tAactivity/.style={thick,draw=red},
Aactivity/.style={very thick,draw=red},
Bactivity/.style={very thick,draw=blue,dashed},
tSactivity/.style={thick,draw=red,postaction={decorate},
decoration={markings,mark=at position .7 with
{\arrow[draw=red]{>}}}},
Sactivity/.style={very thick,draw=red,postaction={decorate},
decoration={markings,mark=at position .7 with
{\arrow[draw=red]{>}}}}
}

\usepackage[colorlinks=true,linkcolor=blue, urlcolor = black, citecolor = blue, anchorcolor = blue]{hyperref}

\usepackage{graphicx}
\usepackage[tight]{subfigure}
\subfiguretopcaptrue
\usepackage{multirow}
\usepackage{subeqnarray}
\usepackage{wasysym}
\usepackage{eufrak}

\newcommand{\ave}[1]{\left\langle #1 \right\rangle}
\newcommand{\Psurv}{P_\text{s}}
\newcommand{\cov}[2]{\text{Cov}\left(#1,#2\right)}

\newcommand{\var}[1]{\text{Var}\left( #1 \right)}
\newcommand{\pgf}{f}
\newcommand{\union}{\cup}
\newcommand{\plaind}{\mathrm{d}}
\newcommand{\bra}[1]{\left\langle#1\right|}
\newcommand{\ket}[1]{\left|#1\right\rangle}

\newcommand{\sandwich}[3]{\left\langle#1\middle|#2\middle|#3\right\rangle}
\newcommand{\creat}{a^\dagger}
\newcommand{\annih}{a}
\newcommand{\atilde}{\widetilde{a}}
\newcommand{\tildephi}{\widetilde\phi}
\newcommand{\creatphi}{\phi^\dagger}
\newcommand{\Eref}[1]{Eq.~(\ref{eq:#1})}
\newcommand{\Esref}[1]{Eqs.~(\ref{eq:#1})}
\newcommand{\eref}[1]{(\ref{eq:#1})}
\newcommand{\sref}[1]{Sec.~\ref{sec:#1}}
\newcommand{\Sref}[1]{Section~\ref{sec:#1}}
\newcommand{\Aref}[1]{Appendix~\ref{sec:#1}}
\newcommand{\elabel}[1]{\label{eq:#1}}
\newcommand{\deltabar}{\delta\mkern-8mu\mathchar'26}
\newcommand{\imag}{\mathring{\imath}}
\newcommand{\dbar}{\plaind\mkern-6mu\mathchar'26}
\newcommand{\dint}[1]{\mathchoice{\!\plaind#1\,}{\!\plaind#1\,}{\!\plaind#1\,}{\!\plaind#1\,}}

\newcommand{\dintbar}[1]{\mathchoice{\!\dbar#1\,}{\!\dbar#1\,}{\!\dbar#1\,}{\!\dbar#1\,}}

\renewcommand{\exp}[1]{\mathchoice{e^{#1}}{\operatorname{exp}\!\left(#1\right)}{\operatorname{exp}\!\left(#1\right)}{\operatorname{exp}\!\left(#1\right)}}
\newcommand{\corresponding}{\hat{=}}

\newcommand{\mA}{\mathcal{A}}
\newcommand{\mD}{\mathcal{D}}
\newcommand{\mG}{\mathcal{G}}
\newcommand{\mO}{\mathcal{O}}
\newcommand{\Num}{N}
\newcommand{\pdf}[2]{\mathcal{P}_{{#1}}\left({#2}\right)}
\newcommand{\mgf}[2]{\mathcal{M}_{{#1}}\left({#2}\right)}
\newcommand{\xtmin}{t_\text{min}}
\newcommand{\tmin}{t_1}

\newcommand{\gtotal}{\hat{g}}
\newcommand{\gbinary}{g}
\newcommand{\Stirling}[2]{\begin{Bmatrix}{#1}\\{#2}\end{Bmatrix}}
\newcommand{\avsh}[2]{V\left(#1,#2\right)}

\newcommand{\cumavsh}{\Big\langle N(t) \Big| N(T)=0\Big\rangle}

\usepackage{xspace}
\newcommand{\latin}[1]{{\it #1}}
\newcommand{\ie}{\latin{i.e.}\@\xspace}

\newcommand{\Erefs}[1]{Eqs.~(\ref{eq:#1})}

\usepackage[tight]{subfigure}
\subfiguretopcaptrue

\usepackage{textcomp} 

\begin{document}
 
\title{Field-theoretic approach to the universality of branching processes}

\author{Rosalba Garcia-Millan}
\email[]{garciamillan16@imperial.ac.uk}
\affiliation{Department of Mathematics, Imperial College London, London SW7 2AZ, United Kingdom}
\author{Johannes Pausch}
\email[]{j.pausch15@imperial.ac.uk}
\affiliation{Department of Mathematics, Imperial College London, London SW7 2AZ, United Kingdom}
\author{Benjamin Walter}
\email[]{b.walter16@imperial.ac.uk}
\affiliation{Department of Mathematics, Imperial College London, London SW7 2AZ, United Kingdom}
\author{Gunnar Pruessner}
\email[]{g.pruessner@imperial.ac.uk}
\affiliation{Department of Mathematics, Imperial College London, London SW7 2AZ, United Kingdom}

\date{6 December 2018}

\begin{abstract}
Branching processes are widely used to model phenomena from networks to neuronal avalanching. In a large class of continuous-time branching processes, we study
 the temporal scaling of the moments of the instant population
size, 
the survival probability, expected avalanche duration, the so-called avalanche 
shape, the $n$-point correlation function and the probability density function 
of the total avalanche size. Previous studies
have shown universality in certain observables of branching processes using probabilistic arguments, however, a comprehensive
description is lacking. We derive the field theory that describes the process and demonstrate how to use it to calculate the
relevant observables and their scaling to leading order in time, revealing the universality of the moments of the population
size. Our results explain why the first and second moment of the offspring distribution are sufficient to fully characterise
the process in the vicinity of criticality, regardless of the underlying offspring distribution. This finding implies that
branching processes are universal. 
We illustrate our analytical results with computer simulations.
\end{abstract}



\maketitle

\section{Introduction \label{intro}}

Branching processes \cite{Harris:1963} are widely used for modelling phenomena in many different subject areas, such as 
avalanches \cite{ZapperiLauritsenStanley:1995,lee2004branching,GleesonDurrett:2017}, 
networks \cite{gilbert1961random,GleesonDurrett:2017,lee2004branching, durrett2006random},
earthquakes \cite{marzocchi2008, AschwandenCorralFontClosCh5:2013}, 
family names \cite{reed2003}, populations of bacteria and cells \cite{kimmel2002, durrett2015branching},
nuclear reactions \cite{williams2013random, pazsit2007neutron},
cultural evolution \cite{rockmore2018} and neuronal avalanches \cite{seshadri2018altered,
BeggsPlenz:2003}.
Because of their mathematical simplicity they play an important role in statistical mechanics \cite{Taeuber:2014} and 
the theory of complex systems \cite{AschwandenCorralFontClosCh5:2013}.

Branching processes are a paradigmatic example of a system displaying a second-order phase transition between extinction (absorbing state)
with probability one and non-zero probability of survival (non-absorbing state) in the infinite time limit.
The critical point in the parameter region at which this transition
occurs is where branching and extinction rates exactly balance,
namely when the expected number of offspring per particle is exactly 
unity \cite{Harris:1963, AschwandenCorralFontClosCh5:2013}.

In the present work we study the continuous-time version of the Galton-Watson branching process \cite{Harris:1963}, 
which is a generalisation
of the birth-death process \cite{Gardiner:1997, GrimmettStirzaker:1992}. In the continuous-time branching process,
particles go extinct or replicate 
into a number of identical offspring at random and with constant Poissonian rates.
Each of the new particles follows the same process. 
The difference between the original Galton-Watson branching process and the continuous-time
branching process we consider here, lies in the waiting times between events. In 
the original Galton-Watson branching process, updates occur in discrete
time steps, while in the continuous-time process we consider, waiting
times follow a Poisson process 
\cite{Gardiner:1997, GrimmettStirzaker:1992}.
However, both processes share many asymptotics \cite{Harris:1963, Taeuber:2014}, and therefore we regard the continuous-time branching process
as the continuum limit of the Galton-Watson branching process.

By using field-theoretic methods, we provide a general framework to determine universal, finite-time scaling properties of 
a wide range of branching processes close to the critical point. The main advantages of this versatile approach are, on the
one hand, the ease with which observables 
are calculated and, on the other hand, the use of
diagrammatic language, which allows us to manipulate the
sometimes cumbersome expressions in a
neat and compact way.
Other methods in the literature developed to study problems related to branching processes, in particular  
relating branching processes to different forms of motion, include the formalism based on the Pal-Bell
equation \cite{pal1958theory,bell1965stochastic,kitamura2019delayed}. 

Moreover, our framework allows us to determine systematically
observables that are otherwise complicated
to manipulate if possible at all.
To illustrate this point, we have calculated in closed form a number of observables
that describe different aspects of the process in the vicinity of the critical point:
we have calculated the moments of the population size as a function of time,
the probability distribution of the population size as a function of time,
the avalanche shape, the two-time and $n$-time correlation functions,
and the total avalanche size and its moments. 

Our results show that branching processes are universal in the vicinity of the critical
point \cite{LeGall:2005, Aldous:1993} in the sense that exactly three quantities (the Poissonian rate and the first
and second moments of the offspring distribution) are sufficient to describe the asymptotics of the
process regardless of the underlying offspring distribution.

The contents of this paper are organised as follows.
In \sref{2FT} we derive the field theory of the continuous-time branching process. 
In \sref{4obs} we use our formalism
to calculate a number of observables in closed form, 
and in \sref{6DC} we discuss our results and our conclusions.
Further details of the calculations can be found in the appendices.

\section{Field Theory of the continuous-time branching process\label{ft}}
\label{sec:2FT} 
\begin{figure}
 \includegraphics[width=\columnwidth]{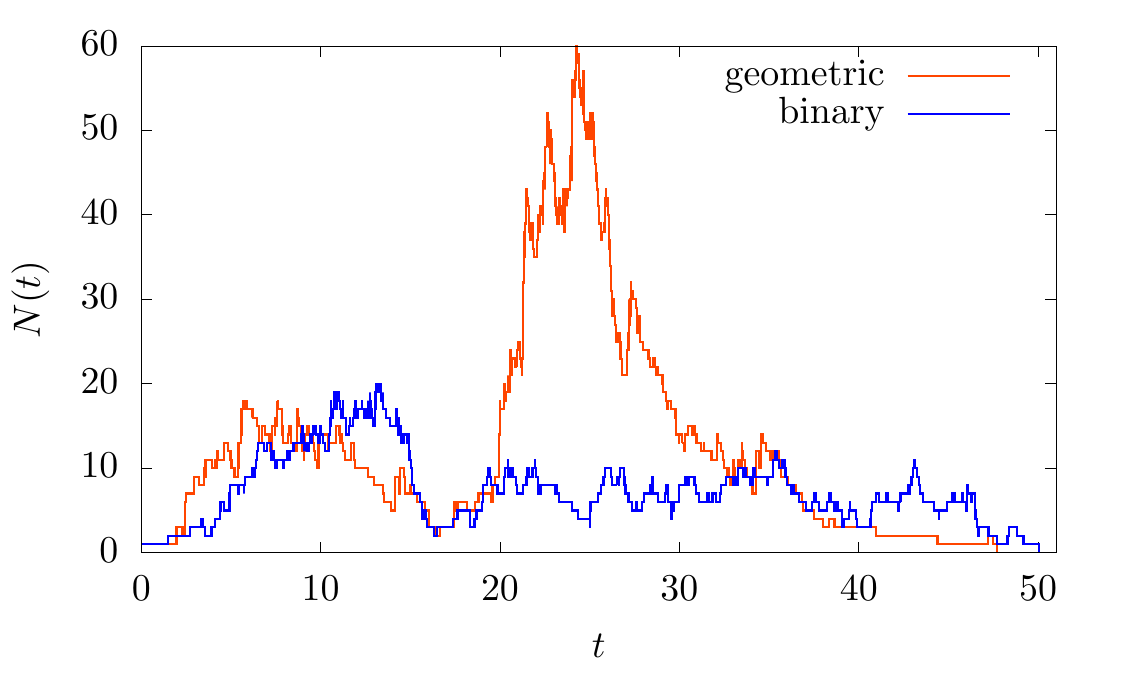}
 \caption{ \label{fig_trajectories} 
 Typical avalanche profiles $N(t)$ of a binary branching process (blue)
 and a branching process with geometric distribution
 of the number of offspring (orange), both at criticality $r=0$,
 and with Poissonian rate $s=1$.}
 \end{figure}

The continuous-time branching process is defined as follows.
We consider a population of $N(t)$ identical particles at time 
$t\geq0$ with initial condition $N(0)=1$. Each particle is
allowed to branch independently into $\kappa$ offspring with Poissonian
rate $s>0$, where $\kappa\in\{0\}\union\mathbb{N}$ is a random variable
with probability distribution $P(\kappa=k)=p_k\in[0,1]$ 
\cite{Gardiner:1997}, Fig.~\ref{fig_trajectories}. In the language of chemical reactions, 
this can be written as the reaction $A\to \kappa A$.

To derive the field theory of this process following the methods by Doi and Peliti \cite{Doi:1976,Peliti:1985,DickmanVidigal:2003,Taeuber:2014}, we first 
write the master equation of the probability $P(\Num,t)$ to find $\Num$ particles at time $t$,
\begin{equation}
\frac{\plaind P(\Num,t)}{\plaind t} 
=  s\sum_k  p_k (\Num-k+1) P(\Num-k+1,t)- s\Num P(\Num,t) ,
\elabel{masterEq}
\end{equation}
with initial condition $P(\Num,0)=\delta_{\Num,1}$. Following work by Doi \cite{Doi:1976}, we cast the master equation in a second quantised form. A system with $\Num$ particles is represented by a Fock-space vector $\ket{\Num}$.  We use the ladder operators $\creat{}$ (creation) and
$a$ (annihilation), which act on $\ket{\Num}$ such that $a\ket{\Num}=N\ket{\Num-1}$ and $\creat{}\ket{\Num}=\ket{\Num+1}$,
and satisfy the commutation relation $[a,\creat{}]=a\creat{}-\creat{}a=1$. 
The probabilistic state of the system is given by\begin{equation}
\ket{\Psi(t)} = \sum_\Num P(\Num,t) \ket{\Num},
\end{equation}
and its time evolution is determined by \Eref{masterEq},
\begin{equation}
\frac{\plaind \ket{\Psi(t)}}{\plaind t} = s\left(\pgf\!\left(\creat{}\right) - \creat{}\right) a \ket{\Psi(t)},
\elabel{Psidot}
\end{equation}
using the probability generating function of $\kappa$,
\begin{equation}
\pgf(z) =\sum_{k=0}^\infty p_k z^k = \ave{z^\kappa},
\elabel{PGF}
\end{equation}
where $\ave{\bullet}$ denotes expectation. 
We define the \emph{mass} $r$ 
as the difference between the extinction and the net branching rates,
\begin{equation}
r = s p_0 - s \sum_{k\geq2}(k-1)p_k = s\left(1-{\langle \kappa \rangle}\right),
\elabel{mass}
\end{equation}
and the rates $q_j$ as 
\begin{equation}
\elabel{def_q_j}
q_j = s\sum_{k} \binom{k}{j} p_k = s\ave{\binom{\kappa}{j}}
=\frac{s}{j!}\pgf^{(j)}(1),
\end{equation}
where $f^{(j)}(1)$ denotes the $j$th derivative of the probability
generating function \Eref{PGF} evaluated at $z=1$.
We assume that the rates $q_j$ are finite.
 In this notation,
the time evolution in \Eref{Psidot} can be 
written as 
\begin{equation}\elabel{formal_op_solution}
\tilde{\mA}\ket{\Psi(t)} = \frac{\plaind}{\plaind{t}} 
\ket{\Psi(t)}
\text{ and thus }
\ket{\Psi(t)} = \exp{\tilde{\mA} t} \ket{\Psi(0)},
\end{equation}
where $\tilde{\mA}$ is the operator
\begin{equation}
\tilde{\mA} =\sum_{j\geq2} q_j \atilde^j a-r\atilde a ,
\elabel{action}
\end{equation}
and $\atilde$ denotes the Doi-shifted creation operator,
$\creat{} = 1+\atilde$.

The sign of the mass $r$, \Eref{mass}, determines in which
regime a particular branching process is in; if $r=0$ the 
process is at the critical point, if $r>0$ the process is in the
subcritical regime and if $r<0$ the process is in the supercritical
regime. Subcritical and critical processes are bound to go extinct
in finite time, whereas supercritical process have a positive
probability of survival \cite{Harris:1963}. 

Following the work by Peliti \cite{Peliti:1985}, \Eref{Psidot} can be cast in path integral form. Here, the creation and annihilation operators $\creat{}$ and $a$ are transformed to time-dependent creation and annihilation fields $\creatphi(t)$ and $\phi(t)$
respectively. Similarly, the Doi-shifted operator $\atilde$ is transformed to the time-dependent Doi-shifted field $\tildephi(t)=\creatphi(t)-1$. The action functional of the resulting field theory is given by
\begin{equation}
\mA\!\left[\tildephi,\phi\right]
\!=\!\!\!\int\limits_{-\infty}^{\infty}\!\!\!\plaind t\left\{
\sum\limits_{j\ge2}q_j\tildephi^j(t)\phi(t)-
\tildephi(t)\left(\frac{d}{dt}\!+\!r\right)\phi(t)\right\}.
\elabel{FTaction}
\end{equation}
Using the Fourier transform 
\newcommand{\phiBoth}{\overset{{\tiny(}\texttildelow{\tiny)}}{\phi}}
\begin{subeqnarray}
\phi(t)&=&\int\dintbar{\omega} \phi(\omega) \exp{-\imag \omega t} 
\quad \text{ with }\quad \dintbar{\omega} = \frac{\dint{\omega}}{2\pi},\\
\phi(\omega)&=&\int\dint{t} \phi(t) \exp{\imag \omega t} ,
\end{subeqnarray}
and identically for $\tildephi(t)$, the action \Eref{FTaction}
becomes local in $\omega$ and the bilinear, \ie the Gaussian part
\begin{equation}
\mA_0\left[\tildephi,\phi\right]=-\int\dintbar{\omega}
\tildephi(-\omega)(-\imag\omega+r)\phi(\omega)
\end{equation}
of the path integral
can be determined in closed form. 
The Gaussian path integral is well-defined only when the
mass is positive, $r>0$.
The non-linear terms, $j\ge2$ in \Eref{FTaction}
are then treated as a perturbation about the Gaussian part,
as commonly done in field theory
\cite{Taeuber:2014,TaeuberHowardVollmayr-Lee:2005}.

 \begin{table*}
\centering
\begin{tabular}{|@{\hspace{10pt}} c @{\hspace{10pt}} c @{\hspace{10pt}} c @{\hspace{10pt}} c @{\hspace{10pt}} c @{\hspace{10pt}} c @{\hspace{10pt}}|}
\hline
\textbf{Observable} & \textbf{Symbol} & \textbf{Equation} & \textbf{Expression} & \textbf{Regime} & \textbf{Figure}\\
\hline
Moments of the number of particles $N(t)$ &$\ave{\Num^n(t)}$ & \eref{momsN}, \eref{momsN_G} & exact asymptote &$r\in\mathbb{R}$&\ref{fig_collapse_moments}\\
Moments of the number of particles $N(t)$ &$\ave{\Num^n(t)}$, $n\in\{1,2,3\}$ & \eref{Nmom1} & exact  &$r\in\mathbb{R}$&\ref{fig_collapse_moments}\\
Moment generating function of $N(t)$ &$\mgf{N}{z}$ &\eref{momsN_mfg} & exact asymptote &$r\in\mathbb{R}$&\\
Probability distribution of $N(t)$ & $P(N,t)$ & \eref{pdf_N} & exact  for binary bp&$r\in\mathbb{R}$&\\
Probability of survival &$\Psurv(t)$& \eref{psurv}, \eref{psurv_crit} & exact  for binary bp &$r\in\mathbb{R}$&\ref{fig_psurv}\\
Distribution of the avalanche duration $T$ &$\pdf{T}{t}$ &\eref{prob-dens-death}, \eref{prob-dens-death0}& exact  for binary bp &$r\geq0$&\\
Expected avalanche duration &$\ave{T}$&\eref{aveT} & exact  for binary bp &$r\geq0$&\ref{fig_extT}\\
Avalanche shape &$V(t,T)$ &\eref{avshape}, \eref{avshape0} & exact  for binary bp &$r\in\mathbb{R}$&\ref{fig_avshape_a}\\
Averaged avalanche shape &$\ave{V(\tau)}_T$&\eref{average_avshape} & exact  for binary bp &$r\in\mathbb{R}$&\ref{fig_avshape_b}\\
Two-point connected correlation function &$\cov{N(t_1)}{N(t_2)}$ &\eref{2pointcf} & exact  &$r\in\mathbb{R}$&\ref{fig_corr}\\
Three-point correlation function &$\ave{N(t_1)N(t_2)N(t_3)}$ &\eref{3ptcorr} &exact  &$r\in\mathbb{R}$&\\\
Moments of the total avalanche size $S$ &$\ave{S^n}$ &\eref{Smoms2}, \eref{Size-Moments}& exact asymptote &$r\geq0$&\\
Moment generating function of $S$ &$\mgf{S}{z}$ &\eref{mgfS}& exact asymptote &$r\geq0$&\\
Distribution of $S$ &$\pdf{S}{x}$ &\eref{pdfS} & exact asymptote &$r\geq0$&\ref{Fig:pdfS}\\
\hline
\end{tabular}
 \caption{\label{tab1} Summary of observables including their equation number and what cases and limits
 are exact. The expressions referred to are, in the limit $r\to0$, either: \emph{exact},
\emph{exact  for binary branching processes} (\ie for other
branching process, our result is the leading order in $q_2/r$ for small $r$ at fixed $rt$),
or provide \emph{exact asymptotes} (that is, our result is the leading order for any kind of branching process). The regime of validity near criticality is either critical and subcritical ($r\geq0$) or all encompassing ($r\in\mathbb{R}$).}
\end{table*}
 
\section{Observables \label{sec:obs}}
\label{sec:4obs}
We use the field theory described above to calculate a number of observables that
have received attention in the literature in various settings. In Table \ref{tab1}
we list all the observables that we have calculated in closed form and the degree of
approximation of our analytical result. Some results are exact for any kind of 
branching process and other results are only exact for binary branching processes.
Those results that are an approximation have the exact asymptotic behaviour.

All observables are constructed on the basis of the probability vector
$\ket{\Psi(t)}$ which evolves according to 
\Eref{formal_op_solution}. If the initial state, $t=0$, consists of a single
particle, then $\ket{\Psi(0)}=\creat{}\ket{0}$ and 
$\ket{\Psi(t)}=\exp{\tilde{\mA} t} \creat{}\ket{0}$. 
Probing the particle number requires the action of the operator
$\creat{}\annih{}$, whose eigenvectors are the pure states $\ket{N}$, 
such that $\creat{}\annih{} \ket{N}=N\ket{N}$. The components of
the vector 
$\creat{}\annih{} \exp{\tilde{\mA} t} \creat{}\ket{0}$ 
are thus the probability that the process has generated $N$ particles 
weighted by $N$. To sum over all states, we further need the projection 
state
\begin{equation}\elabel{projection_state}
\bra{\sun}=\sum\limits_{N=0}^\infty\bra{N}=\sum_{N=0}^\infty \frac{1}{N!} \bra{0}\annih^N = \bra{0}\exp{\annih} \ .
\end{equation}
The expected particle number at time $t$ may thus be written as
\begin{equation}\elabel{example_ave_N}
\ave{N(t)} =
\bra{\sun}
\creat{}\annih{}\, \exp{\tilde{\mA} t} \creat{}
\ket{0} \ .
\end{equation}
More complicated observables and intermediate temporal evolution
can be compiled following the same pattern \cite{TaeuberHowardVollmayr-Lee:2005}. 
To perform
any calculations, the operators need to be normal ordered and 
then mapped to fields as suggested above, 
$\creat{}\to\creatphi(t)=1+\tildephi(t)$
and 
$\annih{}\to\phi(t)$, where the time $t$ corresponds to the
total
time the system has evolved for, \Eref{formal_op_solution}. The
expectation in \Eref{example_ave_N} can thus be written
as 
\begin{equation}
\ave{\Num(t)} = \ave{\creatphi(t)\phi(t)\,\creatphi(0)},
\end{equation}
where $\ave{\mO}$ denotes the path integral
\begin{equation}
\ave{\mO}=\int\mD\left[\tildephi,\phi\right]\mO \exp{\mA\left[\tildephi,\phi\right]}.
\end{equation}

The resulting expressions are most elegantly expressed in 
terms of Feynman diagrams
\cite{Taeuber:2014}. The bare propagator of the field $\phi$
is read off from the bilinear part of the action which, in Fourier space, is
\begin{equation}\elabel{propagator}
\ave{\phi(\omega)\tildephi(\omega')}=\frac{\deltabar(\omega+\omega')}{-\imag\omega+r}\corresponding
\tikz[baseline=-2.5pt]{ \draw[Aactivity] (0.5,0) -- (-0.5,0) node[at end,above] { }; },
\end{equation}
where $\deltabar(\omega+\omega')=2\pi \delta(\omega+\omega')$ denotes
the scaled Dirac-$\delta$ function. Diagrammatically, the bare
propagator is represented by a straight directed line. The directedness
of the propagator reflects the causality (see \Eref{real_time_propagator}) 
of the process in
the time domain as a particle has to be created before it can be 
annihilated but not vice versa.
By convention, in our Feynman diagrams time proceeds
from right to left.

Using the fact that the mass 
$r$ is strictly positive for the Gaussian path integral to converge,
we write the propagator in real time by Fourier transforming, 
\begin{multline}\elabel{real_time_propagator}
\ave{\phi(t)\tildephi(t')}=
\int_{-\infty}^\infty
\dintbar{\omega}
\frac{\deltabar(\omega+\omega')}{-\imag\omega+r}
e^{-\imag\omega t}
e^{-\imag\omega' t'}\\
=
\Theta(t-t')e^{-r(t-t')},
\end{multline}
where $\Theta$ is the Heaviside step function. If $r<0$, the integral in 
\Eref{real_time_propagator} is only convergent for $t<t'$, which violates
causality and therefore yields an unphysical result. For this reason,
we will assume $r>0$ and we will take the limit $r\to0$ where
possible. Therefore, in this paper, the analytical results
obtained through this field theory hold for the critical and subcritical regimes only ($r\geq0$).
However, in some cases we may be able to use probabilistic arguments that
allow us to extend our results to the supercritical case ($r<0$), see \Sref{psurv}.
Furthermore, we will drop
the cumbersome Heaviside $\Theta$ functions, 
assuming suitable choices for the times, such as
$t>t'$ above.

Each of the interaction terms of the form $\tildephi^j\phi$
with $j\ge2$ in the non-linear part of the action 
\Eref{action} come with individual couplings $q_j$, \Eref{def_q_j}.
These are to be expanded perturbatively in. Following
the canonical field theoretic procedure \cite{TaeuberHowardVollmayr-Lee:2005,Taeuber:2014,Peliti:1985}, they are represented by 
(tree-like) amputated vertices such as
\begin{equation}\elabel{vertex_examples}
\tikz[baseline=-2.5pt]{
\draw[Aactivity] (0.5,0) -- (0,0) node[at end,above] {$\,q_2$};
\draw[Aactivity] (130:0.5) -- (0,0);
\draw[Aactivity] (-130:0.5) -- (0,0);},\quad
\tikz[baseline=-2.5pt]{
\draw[Aactivity] (0.5,0) -- (0,0) node[at end,above] {$\,q_3$};
\draw[Aactivity] (130:0.5) -- (0,0);
\draw[Aactivity] (180:0.5) -- (0,0);
\draw[Aactivity] (-130:0.5) -- (0,0);},\quad
\tikz[baseline=-2.5pt]{
\draw[Aactivity] (0.5,0) -- (0,0) node[at end,above] {$\,q_4$};
\draw[Aactivity] (130:0.5) -- (0,0);
\draw[Aactivity] (163:0.5) -- (0,0);
\draw[Aactivity] (-163:0.5) -- (0,0);
\draw[Aactivity] (-130:0.5) -- (0,0);}.
\end{equation}
These vertices are \emph{not} to be confused with the underlying 
branching process, because after the Doi-shift, lines are not 
representative of particles, but of their correlations. For example, the vertex with coupling $q_2$ in \Eref{vertex_examples},
accounts for density-density correlations
due to \emph{any} branching or extinction, 
just like the propagator \Eref{propagator} accounts for all particle density
due to \emph{any} branching or extinction.
After Fourier transforming, these processes
are accounted for regardless of when they 
take place.

The directionality of the diagrams allows us 
to define \emph{incoming} legs and \emph{outgoing} legs of a vertex 
\cite{Taeuber:2014}. 
In the present branching process, all vertices have one incoming leg and 
$j$ outgoing legs. 
We will refer to diagrams that are constructed solely from $q_2$ vertices
as \emph{binary tree diagrams}. The 
most basic such diagram is
$
\tikz[baseline=-2.5pt]{
\draw[Aactivity] (0.3,0) -- (0,0);
\draw[Aactivity] (166:0.3) -- (0,0);
\draw[Aactivity] (-166:0.3) -- (0,0);}
$, which in real time reads
\begin{multline}
\ave{\phi^2(t)
\creatphi(0)
}\corresponding \,2
\tikz[baseline=-2.5pt]{
\draw[Aactivity] (0.5,0) -- (0,0) node[at end,above] {$\,q_2$};
\draw[Aactivity] (130:0.5) -- (0,0);
\draw[Aactivity] (-130:0.5) -- (0,0);}\\
= 2q_2 \!\!\int\!\dintbar{\omega_1}\dintbar{\omega_1'}\dintbar{\omega_2}
\dintbar{\omega_2'}\dintbar{\omega_3}\dintbar{\omega_3'}
\,\exp{-\imag\omega_2t}\exp{-\imag\omega_3t}\\
\times
\deltabar\left(\omega_1\!+\!\omega_2'\!+\!\omega_3'\right)
\frac{\deltabar\left(\omega_1+\omega_1'\right)}{-\imag\omega_1+r}
\frac{\deltabar\left(\omega_2+\omega_2'\right)}{-\imag\omega_2+r}
\frac{\deltabar\left(\omega_3+\omega_3'\right)}{-\imag\omega_3+r}\\
= 2\frac{q_2}{r} \exp{-rt}\left(1-\exp{-rt}\right),
\end{multline}
where the pre-factor $2$ is the combinatorial factor of this
diagram.

The various observables that we calculate in the following are illustrated by numerics for
two different kinds of continuous-time
branching processes. 
Firstly, the binary branching
process with probabilities $p_0,p_2\geq0$ such that $p_0+p_2=1$, 
and secondly, the branching process with geometric 
offspring distribution 
$p_k=p(1-p)^k$ with $p\in[0,1]$.
The mass $r$ \eref{mass} and the rates of the couplings $q_j$ \eref{def_q_j}
are, in each case,
\begin{subeqnarray}{}
&\text{ (binary) }&r_B=s(1-2p_2), \\
&&q_{B2}=sp_2=\frac{s-r_B}{2},\, q_{Bj}=0 \text{ for } j\geq3 ,\nonumber\\
&\text{ (geometric) }&r_G=s\frac{2p-1}{p}, \\
&&q_{Gj}=s\left(\frac{1-p}{p}\right)^j = s\left(1-\frac{r_G}{s}\right)^j .\nonumber
\end{subeqnarray}
Fig.~\ref{fig_trajectories} shows 
typical trajectories for each case.

\subsection{Moments $\ave{\Num^n(t)}$ and their universality}\label{sec:moments}

In the following we will calculate the moments of the number of particles 
$\Num(t)$, which can be determined using the 
particle number operator 
$\creat{}\annih{}$, as introduced above. 
The $n$th moment of
$\Num(t)$ can be expressed as
\begin{eqnarray}
\ave{\Num^n(t)}  &=& \sandwich{\sun}{\left(\creat{}a\right)^n}{\Psi(t)} \nonumber \\
&=&\sum\limits_{\ell=0}^n\Stirling{n}{\ell}\sandwich{\sun}{a^\ell}{\Psi(t)}\nonumber \\
&=&\sum\limits_{\ell=0}^{n}\Stirling{n}{\ell} \ave{\phi^{\ell}(t)\tildephi(0)},
\elabel{momN}
\end{eqnarray}
where $\Stirling{n}{\ell}$ denotes the Stirling number of the second kind
for $\ell$ out of $n$ \footnote{The Stirling numbers of the second kind
can be calculated using the expression
\begin{equation*}
\Stirling{n}{\ell} = \frac{1}{\ell!}\sum_{j=0}^\ell(-1)^{\ell-j}\binom{\ell}{j}j^n.
\end{equation*}
}.
We define the dimensionless function $\gtotal_n(t)$ as the expectation
\begin{equation}\elabel{bubble_vertex}
\gtotal_n(t)=\ave{\phi^n(t)\tildephi(0)}
\corresponding\,
\tikz[baseline=-2.5pt]{
\begin{scope}
  \draw[Aactivity] (-150:0.2) -- (-150:0.8);
  \path [postaction={decorate,decoration={raise=0ex,text along path, text align={center}, text={|\large|....}}}] (170:0.7cm) arc (170:210:0.8cm);
  \draw[Aactivity] (170:0.2) -- (170:0.8);
  \draw[Aactivity] (150:0.2) -- (150:0.8);
\end{scope}
\draw[Aactivity] (0:0.2) -- (0:0.8);
\draw[thick,fill=white] (0,0) circle (0.2cm);
\node at (190:1) {$n$};
} ,
\end{equation}
with $\gtotal_0(t)=\ave{\tildephi(0)} =0$ and 
$\gtotal_1(t) =  \ave{\phi(t)\tildephi(0)} = e^{-rt}$.
The black circle 
in the diagram of \Eref{bubble_vertex}
represents the sum of all possible intermediate nodes, allowing for internal lines.
For instance,
\begin{subequations}{ }
\begin{align}
\gtotal_1(t) 
&\corresponding
\,\tikz[baseline=-2.5pt]{
\draw[Aactivity] (0.8,0) -- (-0.8,0);
\draw[thick,fill=white] (0,0) circle (0.2cm);
} 
=
\tikz[baseline=-2.5pt]{ 
\draw[Aactivity] (0.5,0) -- (-0.5,0) node[at end,above] {}; }\, ,\elabel{g1t}\\
\gtotal_2(t) &\corresponding 
\,\tikz[baseline=-2.5pt]{
\draw[Aactivity] (0.8,0) -- (0,0);
\draw[Aactivity] (150:0.8) -- (0,0);
\draw[Aactivity] (-150:0.8) -- (0,0);
\draw[thick,fill=white] (0,0) circle (0.2cm);}
=
2 \,
\tikz[baseline=-2.5pt]{
\draw[Aactivity] (0.5,0) -- (0,0) node[at end,above] {};
\draw[Aactivity] (130:0.5) -- (0,0);
\draw[Aactivity] (-130:0.5) -- (0,0);}\, ,\elabel{g2t}\\
\gtotal_3(t) &\corresponding
\,\tikz[baseline=-2.5pt]{
\draw[Aactivity] (0.8,0) -- (0,0);
\draw[Aactivity] (150:0.8) -- (0,0);
\draw[Aactivity] (180:0.8) -- (0,0);
\draw[Aactivity] (-150:0.8) -- (0,0);
\draw[thick,fill=white] (0,0) circle (0.2cm);}
= 6\,
\,\tikz[baseline=-2.5pt]{
\draw[Aactivity] (0.5,0) -- (0,0) node[at end,above] {};
\draw[Aactivity] (130:0.5) -- (0,0);
\draw[Aactivity] (180:0.5) -- (0,0);
\draw[Aactivity] (-130:0.5) -- (0,0);}
+ 12\,
\tikz[baseline=-2.5pt]{
\draw[Aactivity] (1,0) -- (0,0) node [at end,above] {};
\draw[Aactivity] (130:0.5) -- (0,0);
\draw[Aactivity] (-130:0.5) -- (0,0);
\node (a) at (0.5,0) {}; 
\draw[Aactivity] (a)+(-130:0.5) -- (0.5,0) node [at end,above] {};}\, 
,\elabel{g3t}\\
\gtotal_4(t) &\corresponding 
\,\tikz[baseline=-2.5pt]{
\draw[Aactivity] (0.8,0) -- (0,0);
\draw[Aactivity] (150:0.8) -- (0,0);
\draw[Aactivity] (170:0.8) -- (0,0);
\draw[Aactivity] (-170:0.8) -- (0,0);
\draw[Aactivity] (-150:0.8) -- (0,0);
\draw[thick,fill=white] (0,0) circle (0.2cm);}
= 24\,
\tikz[baseline=-2.5pt]{
\draw[Aactivity] (0.5,0) -- (0,0) node[at end,above] {};
\draw[Aactivity] (130:0.5) -- (0,0);
\draw[Aactivity] (163:0.5) -- (0,0);
\draw[Aactivity] (-163:0.5) -- (0,0);
\draw[Aactivity] (-130:0.5) -- (0,0);}
+ 48\,
\tikz[baseline=-2.5pt]{
\draw[Aactivity] (1,0) -- (0,0) node [at end,above] {};
\draw[Aactivity] (130:0.5) -- (0,0);
\draw[Aactivity] (180:0.5) -- (0,0);
\draw[Aactivity] (-130:0.5) -- (0,0);
\node (a) at (0.5,0) {};
\draw[Aactivity] (a)+(-130:0.5) -- (0.5,0);}\\
&+ 72\,
\tikz[baseline=-2.5pt]{
\draw[Aactivity] (1,0) -- (0,0) node [at end,above] {};
\draw[Aactivity] (130:0.5) -- (0,0);
\draw[Aactivity] (-130:0.5) -- (0,0);
\node (a) at (0.5,0) {};
\draw[Aactivity] (a)+(-130:0.5) -- (0.5,0);
\draw[Aactivity] (a)+(130:0.5) -- (0.5,0);}
+ 24\,
\tikz[baseline=-2.5pt]{
\draw[Aactivity] (0.5,0) -- (0,0) node[at end,above] {};
\node (a) at (-130:0.4) {};
\node (b) at (130:0.4) {};
\draw[Aactivity] (130:0.8) -- (0,0);
\draw[Aactivity] (-130:0.8) -- (0,0);
\draw[Aactivity] (-130:0.4)+(150:0.3) -- (-130:0.5);
\draw[Aactivity] (130:0.4)+(-150:0.3) -- (130:0.5);}
+ 96\,
\tikz[baseline=-2.5pt]{
\draw[Aactivity] (1.5,0) -- (0,0) node [at end,above] {};
\draw[Aactivity] (130:0.5) -- (0,0);
\draw[Aactivity] (-130:0.5) -- (0,0);
\node (a) at (0.5,0) {};
\draw[Aactivity] (a)+(-130:0.5) -- (0.5,0);
\node (b) at (1,0) {};
\draw[Aactivity] (b)+(-130:0.5) -- (1,0);\elabel{g4t}}\, ,\nonumber
\end{align}
\elabel{ex_gtotal_diagrams}
\end{subequations}
\noindent where the coefficient in front of each diagram is its symmetry factor,
which is included in the representation involving the black circle, 
\Eref{bubble_vertex}.

The tree diagrams follow a pattern, whereby 
$\gtotal_n$ involves all $\gtotal_m$ with $m<n$. 
For $n\ge2$ this can be expressed as the 
recurrence relation
\begin{multline}{ }
\gtotal_n(t)
\corresponding\!\!
\sum_{k=2}^n \sum_{m_1,\ldots,m_k=1}\!\!\binom{n}{m_1,\ldots,m_k}\!\!
\tikz[baseline=-2.5pt,scale=0.8]{
\draw[Aactivity] (0.5,0) -- (0,0);
  \path [postaction={decorate,decoration={raise=0ex,text along path, text align={center}, text={|\large|......}}}] (170:0.7cm) arc (170:220:0.8cm);
  \node at (190:0.9) {$k$};
  \node (a) at (170:1.5) {};
  \path [postaction={decorate,decoration={raise=0ex,text along path, text align={center}, text={|\large|....}}}] (a)+(170:0.7cm) arc (170:210:0.8cm);
  \draw[Aactivity] (a)+(-150:0.2) -- ++(-150:0.8);
  \draw[Aactivity] (a)+(170:0.2) -- ++(170:0.8);
  \draw[Aactivity] (a)+(150:0.2) -- ++(150:0.8);
\draw[Aactivity] (a)+(0:0.2) -- (0,0);
\draw[thick,fill=white] (a)+(0,0) circle (0.2cm);
\node at (178:2.5) {$m_2$};
\begin{scope}
  \node (b) at (140:1.75) {};
  \path [postaction={decorate,decoration={raise=0ex,text along path, text align={center}, text={|\large|....}}}] (b)+(170:0.7cm) arc (170:210:0.8cm);
  \draw[Aactivity] (b)+(-150:0.2) -- ++(-150:0.8);
  \draw[Aactivity] (b)+(170:0.2) -- ++(170:0.8);
  \draw[Aactivity] (b)+(150:0.2) -- ++(150:0.8);
\end{scope}
\draw[Aactivity] (b)+(0:0.2) -- (0,0);
\draw[thick,fill=white] (b)+(0,0) circle (0.2cm);
\node at (160:2.5) {$m_1$};
\begin{scope}
  \node (c) at (-140:1.75) {};
  \path [postaction={decorate,decoration={raise=0ex,text along path, text align={center}, text={|\large|....}}}] (c)+(170:0.7cm) arc (170:210:0.8cm);
  \draw[Aactivity] (c)+(-150:0.2) -- ++(-150:0.8);
  \draw[Aactivity] (c)+(170:0.2) -- ++(170:0.8);
  \draw[Aactivity] (c)+(150:0.2) -- ++(150:0.8);
\end{scope}
\draw[Aactivity] (c)+(0:0.2) -- (0,0);
\draw[thick,fill=white] (c)+(0,0) circle (0.2cm);
\node at (207:2.6) {$m_k$};
},
\end{multline}
where $\binom{n}{m_1,\ldots,m_k}$ denotes the multinomial coefficient
with the implicit constraint of $m_1+\ldots+m_k=n$.
Including $\gtotal_1(t)$ from \Erefs{g1t} and \eref{real_time_propagator}, this may be written as
\begin{multline}\elabel{gnrec}
\gtotal_n(t) = 
\delta_{n,1}\exp{-rt}
+\Biggl(\sum_{k=2}^n q_k \sum_{m_1,\ldots,m_k}\binom{n}{m_1,\ldots,m_k}\\
\times \int_0^t \dint{t'} \exp{-r(t-t')}\gtotal_{m_1}(t')\gtotal_{m_2}(t')\cdots \gtotal_{m_k}(t')\Biggr)
\ ,
\end{multline}
where the integral accounts for the propagation up until time $t-t'\in[0,t]$
when a branching into (at least) $k$ particles takes place, each of which will
branch into (at least) $m_k$ particles at some later time within $[t-t',t]$.

We proceed by determining the leading order behaviour of $\gtotal_n$ in small $r$, starting
with a dimensional argument. Since 
\begin{equation}
\ave{\Num^n(t)}
=\sum\limits_{\ell=0}^{n}\Stirling{n}{\ell} \gtotal_\ell(t)
\elabel{Ngtotal}
\end{equation}
from \Erefs{momN} and \eref{bubble_vertex}, 
$\ave{\Num^n(t)}$ being dimensionless implies the same for $\gtotal_n(t)$. Our notation for
the latter obscures the fact that $\gtotal_n(t)$ is also a function of $r$ and all $q_j$, which,
by virtue of $s$, are rates and thus have the inverse dimension of $t$. We may therefore write
\newcommand{\gbar}{\overline{g}}
\newcommand{\qbar}{\overline{q}}
\begin{equation}
\gtotal_n(t) = \gbar_n(rt;\qbar_2,\qbar_3,\ldots)
\elabel{gbar_def}
\end{equation}
where $\qbar_j=q_j/r$ are dimensionless couplings. 
Dividing $q_j$ by \emph{any} rate renders the result dimensionless, but only the particular
choice of dividing by $r$ ensures that all couplings only ever enter multiplicatively 
(and never as an inverse), thereby enabling us to extract the asymptote of $\gtotal_n(t)$ in the
limit of small $r$, as we will see in the following. 

Writing \Eref{gnrec} as
\begin{eqnarray}\elabel{gbarnrec}
&&\gbar_n(y;\qbar_2,\qbar_3,\ldots) = 
\delta_{n,1}\exp{-y}\\
&&\quad+\!\Biggl(\sum_{k=2}^n \qbar_k\!\!\! \sum_{m_1,\ldots,m_k}\!\!\!\binom{n}{m_1,\ldots,m_k}
 \int_0^y\!\! \dint{y'}\exp{-(y-y')}\nonumber\\
&&\qquad\times
\gbar_{m_1}\!(y';\qbar_2,\ldots)
\gbar_{m_2}\!(y';\ldots)\cdots
\gbar_{m_k}\!(y';\ldots)\!\!\Biggr)
\ ,\nonumber
\end{eqnarray}
the dominant terms in small $r$ and fixed $y=rt$ are those that contain products
involving the largest number of factors of $\qbar_j\propto r^{-1}$.
Since each $\qbar_j$ corresponds to a branching, diagrammatically these terms are those
that contain the largest number of vertices, \ie those that are entirely made up of
binary branching vertices $q_2$. This coupling,
$q_2=\ave{\kappa(\kappa-1)}\!/2$ ,
cannot possibly vanish if there is any branching taking place at all.
From \Erefs{gbar_def} and \eref{gbarnrec} it follows that $\gtotal_n(t)\propto (q_2/r)^{(n-1)}$
to leading order in small $r$ at fixed $y=rt$. Terms of that order 
are due to binary tree diagrams, whose contribution we denote by $g_n(t)$ in the following.
For instance, $\gbinary_1(t)=\gtotal_1(t)$, $\gbinary_2(t)=\gtotal_2(t)$,
\begin{subeqnarray}
\gbinary_3(t) &\corresponding& 
12\,
\tikz[baseline=-2.5pt]{
\draw[Aactivity] (1,0) -- (0,0) node [at end,above] {};
\draw[Aactivity] (130:0.5) -- (0,0);
\draw[Aactivity] (-130:0.5) -- (0,0);
\node (a) at (0.5,0) {};
\draw[Aactivity] (a)+(-130:0.5) -- (0.5,0);},\\
\gbinary_4(t)
&\corresponding&
24\,
\tikz[baseline=-2.5pt]{
\draw[Aactivity] (0.5,0) -- (0,0) node[at end,above] {};
\node (a) at (-130:0.4) {};
\node (b) at (130:0.4) {};
\draw[Aactivity] (130:0.8) -- (0,0);
\draw[Aactivity] (-130:0.8) -- (0,0);
\draw[Aactivity] (-130:0.4)+(150:0.3) -- (-130:0.5);
\draw[Aactivity] (130:0.4)+(-150:0.3) -- (130:0.5);}
+ 96\,
\tikz[baseline=-2.5pt]{
\draw[Aactivity] (1.5,0) -- (0,0) node [at end,above] {};
\draw[Aactivity] (130:0.5) -- (0,0);
\draw[Aactivity] (-130:0.5) -- (0,0);
\node (a) at (0.5,0) {};
\draw[Aactivity] (a)+(-130:0.5) -- (0.5,0);
\node (b) at (1,0) {};
\draw[Aactivity] (b)+(-130:0.5) -- (1,0);}.
\end{subeqnarray}
To summarise,
\emph{$\gtotal_n(t)$ is dominated by those terms that correspond 
to binary tree diagrams}, which are the trees $g_n$ that have the largest
number of vertices for any fixed $n$, \ie
\begin{equation}
\ave{\phi^n(t)\tildephi(0)} = \gtotal_n(t) = \gbinary_n(t)
+\mO\left(\left(1-{\langle \kappa \rangle}\right)^{-(n-2)}\right),
\elabel{gbin_approx_gtot}
\end{equation}
where the correction in fact vanishes for $n<3$.

As far as the asymptote in small $r$ is concerned, we may thus replace
$\gtotal_\ell$ in \Eref{Ngtotal} by $\gbinary_\ell$. Among the
$\gtotal_\ell\sim r^{-(\ell-1)}$ with $\ell=0,1,,\dots,n$, the dominating
term is $\gbinary_n$ so that the $n$th moment of the particle number
$\Num$ is, to leading order,
\begin{equation}
\ave{N^n(t)}\simeq\gbinary_n(t),
\elabel{NnMoment}
\end{equation}
although exact results, as shown in \Eref{Nmom1}, are easily derived
using \Erefs{momN}, \eref{bubble_vertex}, \eref{ex_gtotal_diagrams},
and \eref{gnrec}. On the basis of \eref{gnrec}
the recurrence relation of $\gbinary_n$ is give by
\begin{equation}
\gbinary_n(t)=\delta_{n,1}\exp{-rt}
+ q_2 \sum_{m=1}^{n-1}\binom{n}{m}
\int_0^t \plaind t'\exp{-r(t-t')}\gbinary_{m}(t')\gbinary_{n-m}(t'),
\end{equation}
whose exact solution is
\begin{equation}
\gbinary_n(t) = n!\exp{-rt}\left(\frac{q_2}{r}\left(1-\exp{-rt}\right)\right)^{n-1}.
\elabel{gnt_sol}
\end{equation}

We draw two main conclusions from our results. 
Firstly, that near the critical
point $r=0$, the branching process is solely 
characterised by the two parameters
$r$ and $q_2$. We therefore conclude that this process displays \emph{universality},
in the sense that its asymptotia is exactly the same for any
given $r$ and $q_2$ regardless of the underlying offspring distribution.
In particular, certain ratios of the moments of the particle number are
universal constants (they do not depend on any parameters nor variables).
For $k,\ell\in\mathbb{N}$ and $m\in{0,\dots,k-1}$, we find the constant ratios
\begin{equation}\elabel{TimeMomentRatio}
\frac{\ave{N^k(t)}\ave{N^\ell(t)}}{\ave{N^{k-m}(t)}\ave{N^{\ell+m}}}=\frac{k!\ell!}{(k-m)!(\ell+m)!}.
\end{equation}

\begin{figure}
 \includegraphics[width=\columnwidth]{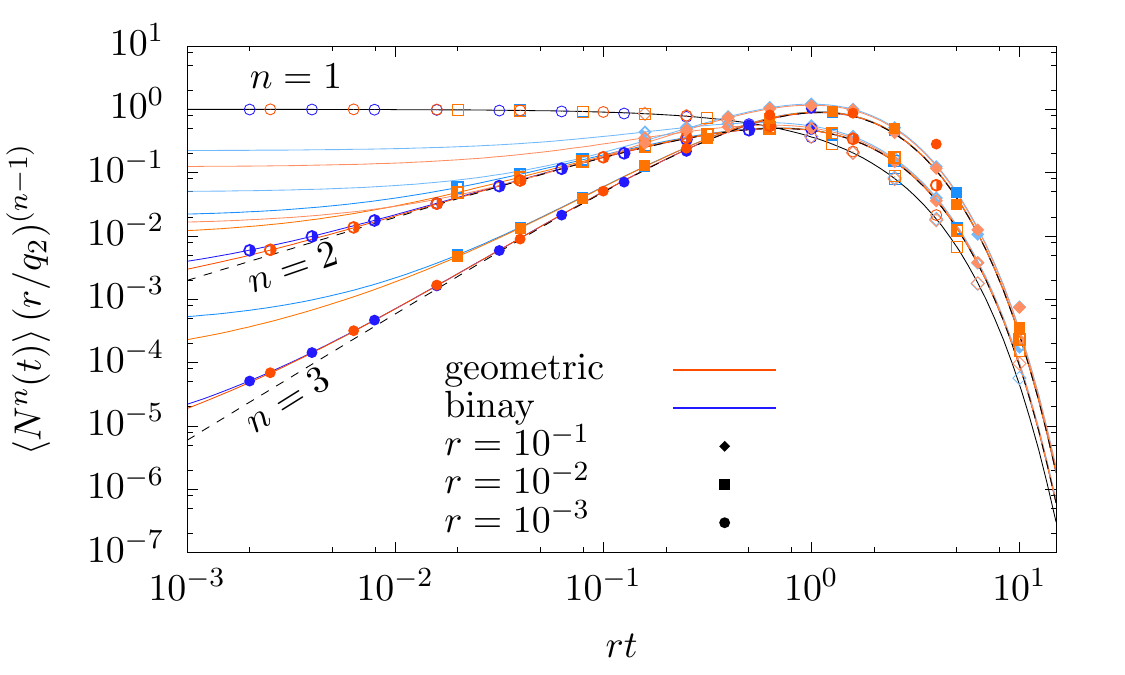}
 \caption{ \label{fig_collapse_moments} 
 Data collapse of the moments $\ave{N(t)}$, $\ave{N^2(t)}$ and $\ave{N^3(t)}$,
 as a function of rescaled time $rt$ as of 
 \Eref{momsN}. Symbols show results for the binary branching process (blue)
 and the branching process with geometric distribution of offspring (orange),
 both with $r\in\{10^{-3}, 10^{-2},10^{-1}\}$ and $s=1$. 
 Solid lines indicate the exact solution in \Eref{Nmom1} and dashed lines
 indicate our approximation in \Eref{NnMoment}.
}
 \end{figure}

Secondly, our results show that the scaling form of the moments is
\begin{equation}
\ave{\Num^n(t)} \simeq ({q_2}{t})^{n-1}\mG_n(rt),
\elabel{momsN}
\end{equation}
where $\mG_n$ is the scaling function 
\begin{equation}
\mG_n(y) = n!\exp{-y}\left(\frac{1-e^{-y}}{y}\right)^{n-1},
\elabel{momsN_G}
\end{equation}
and the argument $y=rt$ is the rescaled time, 
Fig.~\ref{fig_collapse_moments}.
The asymptotes of $\mG_n(y)$ characterise the behaviour
of the branching process in each regime,
\begin{equation}
\mG_n(y) \simeq \left\{
\begin{array}{l l}
n! & \text{ for }y=0,\\
n! \,y^{-(n-1)}e^{-y} & \text{ for }y\to\infty.
\end{array}
\right.
\end{equation}
Moreover, from \Eref{momsN}, we find that
the moment generating function $\mgf{N}{z} = \ave{e^{Nz}}$ is
\begin{equation}
\mgf{N}{z} \simeq 1+\frac{ze^{-rt}}{1+z\frac{q_2}{r}\left(e^{-rt}-1\right)}.
\elabel{momsN_mfg}
\end{equation}

 \subsection{Probability distribution of $N(t)$, probability of survival $\Psurv(t)$ and expected avalanche duration $\ave{T}$}\label{sec:psurv}
 Using \Eref{momN} and the identity  \footnote{
 The Stirling numbers of the second kind satisfy the identity
$$  N^n = \sum\limits_{\ell=0}^{n}\Stirling{n}{\ell} N(N-1)\ldots(N-\ell+1) . $$}
of Stirling numbers of the second kind,
 we deduce that the falling factorial moments of $N(t)$ are
 \begin{equation}
\ave{\phi^{\ell}(t)\tildephi(0)} = \ave{N(t)(N(t)-1)\ldots(N(t)-\ell+1)}.
 \end{equation}
 Therefore, the probability generating function of $N(t)$ is
 \begin{eqnarray}
 \pdf{N(t)}{z} 
 &=& \sum_{\ell=0}^\infty \ave{N(t)(N(t)-1)\ldots(N(t)-\ell+1)}\frac{(z-1)^\ell}{\ell!}\nonumber\\
 &=& \sum_{\ell=0}^\infty \ave{\phi^{\ell}(t)\tildephi(0)}\frac{(z-1)^\ell}{\ell!},
 \end{eqnarray}
 and the probability distribution of $N(t)$ is, using \Erefs{gbin_approx_gtot} and \eref{gnt_sol},
 \begin{subeqnarray}
 \elabel{pdf_N}
 P(N,t) &=& \frac{1}{N!} \frac{\plaind^N}{\plaind z^N}\left(\pdf{N(t)}{z}\right)\big|_{z=0} \\
 &\simeq& \sum_{\ell=N}^\infty \binom{\ell}{N}\frac{(-1)^{\ell-N}}{\ell!}g_\ell(t)\\
 &=& 
 \left\{
\begin{array}{l l}
1-\frac{e^{-rt}}{1+\frac{q_2}{r}\left(1-e^{-rt}\right)} & \text{ if }N=0,\\
\frac{e^{-rt}\left(\frac{q_2}{r}\left(1-e^{-rt}\right)\right)^{N-1}}{\left(1+\frac{q_2}{r}\left(1-e^{-rt}\right)\right)^{N+1}} & \text{ if }N>0,
\elabel{sol_PNt}
\end{array}
\right.
 \end{subeqnarray}
which satisfies the initial condition $P(\Num,0)=\delta_{\Num,1}$ and is an exact result for binary branching
processes, consistent with \cite{pazsit2007neutron}.

It is straightforward to check that \Eref{sol_PNt} satisfies the master equation \eref{masterEq} and the initial condition
in the binary branching case. Due to the uniqueness of solutions of a system of coupled linear ordinary differential
equations, the solution in \Eref{sol_PNt} is the only solution. In particular, this solution holds in the supercritical 
case, $r<0$. Reconstructing back the path that has lead us here, we find that $g_\ell(t)$ \emph{is} the $\ell$th falling
factorial moment of $N(t)$, $\ave{N(t)(N(t)-1)\ldots(N(t)-\ell+1)}$, for binary branching processes including the
supercritical case and, therefore, most expressions derived from $g_\ell(t)$ can be extended to $r<0$.
In what follows, we will specify which expressions hold in the supercritical case.

\begin{figure}
 \includegraphics[width=0.49\textwidth]{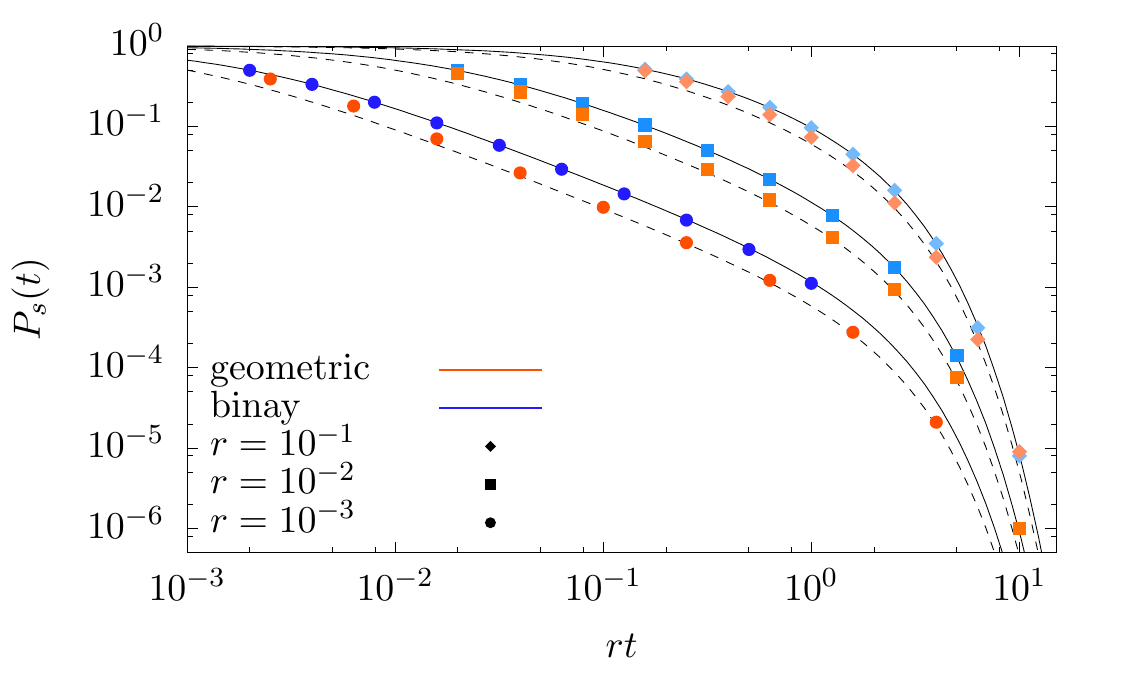}
 \caption{ \label{fig_psurv} 
 Probability of survival as a function of rescaled time $rt$ as of \Eref{psurv}.
 Symbols show numerical results for the binary branching process (blue)
 and the branching process with geometric distribution of offspring (orange),
 both with $r\in\{10^{-3}, 10^{-2}, 10^{-1}\}$ and $s=1$. 
 Lines indicate the result in \Eref{psurv}, which is exact for binary 
 branching (solid lines) and approximate otherwise (dashed lines).
 As $r$ gets closer to the critical value, $r=0$, the curves $\Psurv(t)$ 
 flatten and resemble the power law in \Eref{psurv_crit}, which 
 has exponent $-1$.
}
 \end{figure}

The probability of survival $\Psurv(t)$ is the probability that there is at least
one particle at time $t$, \ie $\Psurv(t)=P(\Num(t)\geq1)$. 
Therefore, from \Eref{sol_PNt}, 
\begin{equation}
\Psurv(t) = 1-P(0,t) = \frac{e^{-rt}}{1+\frac{q_2}{r}\left(1-e^{-rt}\right)},
\elabel{psurv}
\end{equation}
and at the critical point,
\begin{equation}
\lim\limits_{r\rightarrow0}\Psurv(t)\simeq\frac{1}{1+q_2 t},
\elabel{psurv_crit}
\end{equation}
which is consistent with 
\citep{Garcia-MillanFont-ClosCorral:2016,
WeiPruessner:2016,corral2016exact}, Fig.~\ref{fig_psurv}.

\begin{figure}
\includegraphics[width=\columnwidth]{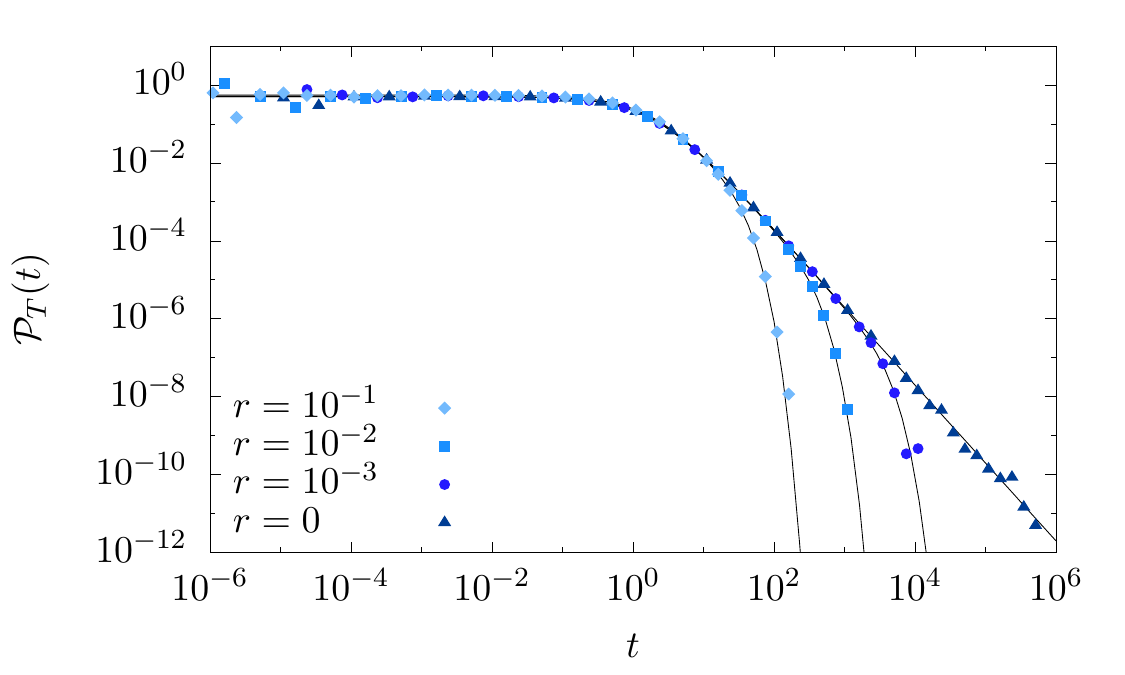}
\caption{\label{fig_extT} Probability density function of the avalanche
duration $\pdf{T}{t}$ for the binary branching process
with $r\in\{0,10^{-3}, 10^{-2}, 10^{-1}\}$ and $s=1$. Solid lines
represent our result in \Esref{prob-dens-death} and \eref{prob-dens-death0},
which is exact for binary branching. Symbols show numerical results.}
\end{figure}

We define the avalanche duration $T$ as the exact time where an
avalanche dies, \ie the time $t$ when the process reaches the 
absorbing state, $T=\text{min}\{t|\Num(t)=0\}$.
The probability of survival $\Psurv(t)$ gives the probability
that $T>t$. Thus, $1-P_s(t)$ is the cumulative distribution function of the 
time of death and its probability density function is
\begin{equation}
\elabel{prob-dens-death}
\pdf{T}{t}=-\frac{d\Psurv(t)}{dt}\simeq\frac{re^{rt}\left(1+\frac{q_2}{r}\right)}{\left(\frac{q_2}{r}-e^{rt}\left(1+\frac{q_2}{r}\right)\right)^2},
\end{equation}
and at the critical point,
\begin{equation}
\elabel{prob-dens-death0}
\lim_{r\to0}\pdf{T}{t}\simeq\frac{q_2}{(1+q_2t)^2},
\end{equation}
see Fig.~\ref{fig_extT}.
It follows from \eref{prob-dens-death} that the expected avalanche duration is
\begin{equation}
\ave{T}\simeq\frac{1}{q_2}\log\left(1+\frac{q_2}{r}\right).
\elabel{aveT}
\end{equation}
Because the derivation of \Eref{prob-dens-death} relies on a finite termination time, we cannot assume that it remains 
valid in the supercritical case, and similarly for \eref{aveT}.


\subsection{Avalanche shape $\avsh{t}{T}$ \label{sec:avsh}}

The avalanche shape $\avsh{t}{T}$ is defined
as the average of the temporal profiles $N(t)$ conditioned
to extinction at time $T$ \cite{GleesonDurrett:2017,DobrinevskiETAL:2015,SheikhETAL:2016,
PapanikolaouETAL:2011, BaldassarriETAL:2003,WillisPruessner:2018,
LaursonETAL:2013}.
Closed form expressions of the avalanche
shape have been calculated in other models such as
avalanches in elastic interfaces \cite{DobrinevskiETAL:2015},
the Barkhausen noise \cite{PapanikolaouETAL:2011},
the discrete-time Ornstein-Uhlenbeck process
\cite{BaldassarriETAL:2003}.
An implicit expression of avalanche shape of branching processes
is given in \cite{GleesonDurrett:2017}.

To produce an explicit expression we
first calculate the expected number of particles at time $t$
of a branching process conditioned to being extinct by time $T$, $\cumavsh$.
In terms of ladder operators,
\begin{equation}
\cumavsh=\sandwich{0}{\exp{\hat{\mA}(T-t)}\creat\annih\exp{\hat{\mA}t}\creat}{0},
\elabel{defshape0}
\end{equation}
which means that a particle is created from the vacuum,
the system is allowed to evolve for time $t$, the number of particles
is measured, and the system evolves further for time $T-t$. 
Finally, all possible trajectories are "sieved" so that
only those avalanches are whose number of particles is $0$ at time $T$
taken into account.
The path integral expression of \Eref{defshape0} is
\begin{eqnarray}{}
&&\cumavsh=\ave{\exp{-\phi(T)}\phi^\dagger(t)\phi(t)\phi^\dagger(0)}\nonumber\\
&&=\ave{\phi(t)\tildephi(0)}+ \sum_{n\geq1}\frac{(-1)^n}{n!} 
\\&&\quad \times\left(\ave{\phi^n(T)\phi(t)\tildephi(0)}\nonumber
    +\ave{\phi^n(T)\tildephi(t)\phi(t)\tildephi(0)}\right). \nonumber
\end{eqnarray}
The two terms in the bracket have asymptotes
\begin{eqnarray}{}
&&\ave{\phi^n(T)\phi(t)\tildephi(0)}
\simeq
\sum_{k=1}^n\sum_{m_1,\ldots,m_k}\binom{n}{m_1,\ldots,m_k}\nonumber\\
&&\qquad\times\frac{1}{k!}\,
g_{m_1}(T-t)\cdots g_{m_k}(T-t)g_{k+1}(t),
\elabel{IIn}
\end{eqnarray}
and
\begin{eqnarray}{}
&&\ave{\phi^n(T)\tildephi(t)\phi(t)\tildephi(0)}
\simeq\sum_{k=1}^n\sum_{m_1,\ldots,m_k}
\binom{n}{m_1,\ldots,m_k}\nonumber\\
&&\qquad\times\frac{1}{(k-1)!}\,
g_{m_1}(T-t)\cdots g_{m_k}(T-t)g_{k}(t),
\elabel{In}
\end{eqnarray}
with the constraint $m_1+\ldots+m_k=n$ in both cases. 
Both expressions are exact in case of binary branching. Their diagrammatic
representation and closed form expressions can be found in \Aref{diagrams}.
Using the expression of 
$g_n(t)$ in \Eref{gnt_sol} and the number of 
combinations of $n$ legs into $k$ groups, we have
\begin{eqnarray}
&&\cumavsh \\
&&\quad = \exp{-rt} - \Psurv(T) \left[ 1+\frac{q_2}{r}\left(1-\exp{-rt}\right)\left(2-\frac{\Psurv(T)}{\Psurv(t)}\right)\right],\nonumber
\end{eqnarray}
where $\Psurv(t)$ is given in \Eref{psurv}.

\begin{figure}
\subfigure[\hspace{230pt}]{ \label{fig_avshape_a} \includegraphics[width=\columnwidth]{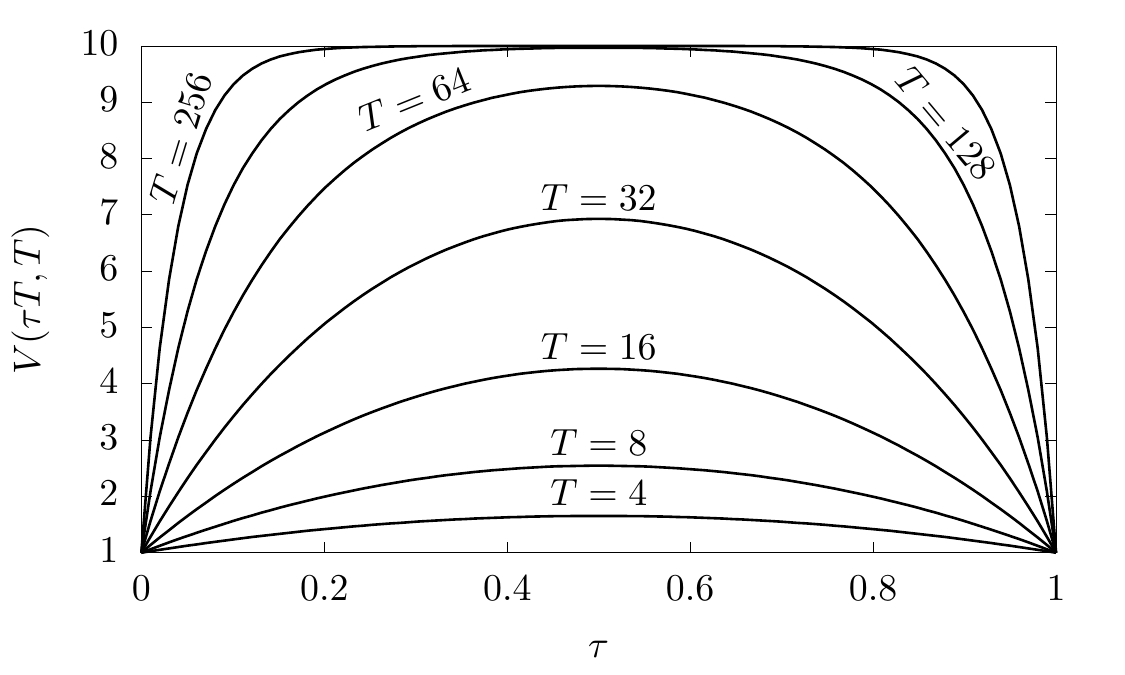}}
\subfigure[\hspace{230pt}]{ \label{fig_avshape_b}\includegraphics[width=\columnwidth]{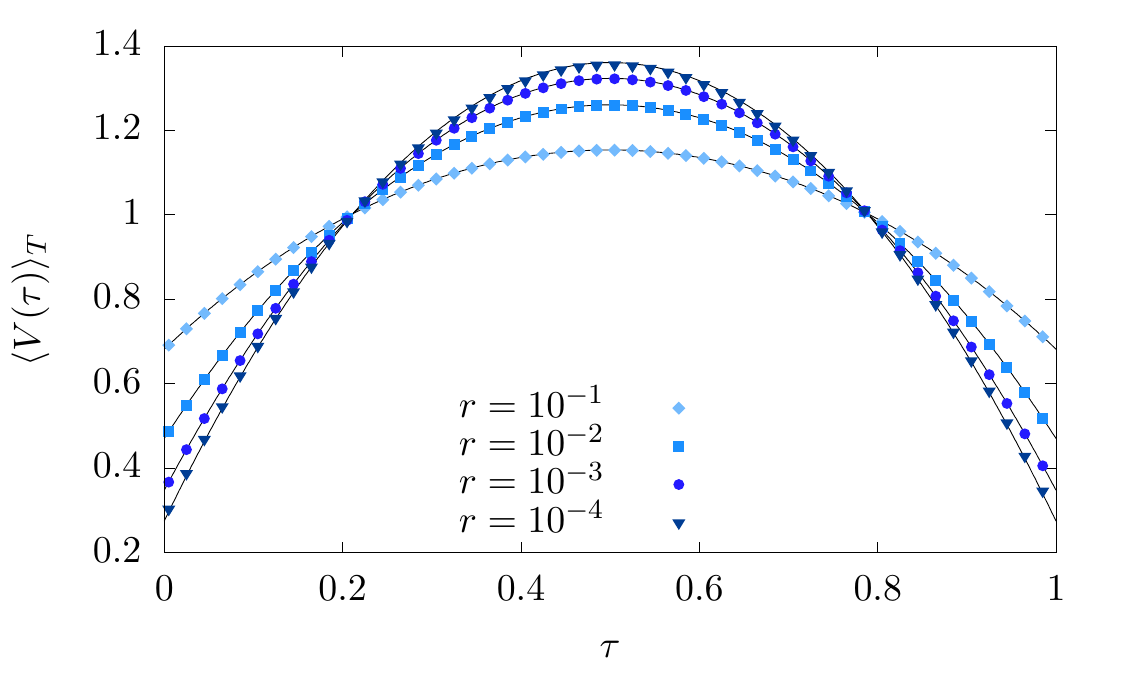}}
\caption{\label{fig_avshape}
In \subref{fig_avshape_a}, avalanche shape $V$
with rescaled time $\tau=t/T$ for different times of extinction $T$, 
$r=10^{-1}$ and $q_2=0.45$ as of \Eref{avshape}.
The shapes are symmetric and flatten
as $T$ increases with the upper bound given in \Eref{avshape_bound}. 
However, this observable is numerically unaccessible because it is computationally 
unfeasible to obtain a large enough sample of avalanches in the subcritical regime
conditioned to extinction at large times. Instead, in \subref{fig_avshape_b} we
show an observable that is accessible both numerically and analytically, the 
averaged avalanche shape $\ave{V(\tau)}_T$, that is for each $r$, avalanches are
rescaled in time to the interval $[0,1]$, their shapes are averaged and
normalised regardless of their extinction times $T$. Numerical results shown as symbols are
for the binary branching process with $r\in\{10^{-4},10^{-3}, 10^{-2}, 10^{-1}\}$ and $s=1$,
and are in agreement with \Eref{average_avshape} (solid lines), which is an exact expression for binary
branching processes. 
We find that the shape tends to a parabola as $r$ approaches the 
critical point.
}
\end{figure}

In order to account solely for those instances that become extinct exactly at 
time $T$, the expectation $\cumavsh$ is to be
differentiated with respect to $T$, and in order to account for the factor
due to conditioning to extinction, we need to divide the result by
$-\frac{\plaind}{\plaind t}\Psurv(t)$, yielding,
\begin{eqnarray}
\avsh{t}{T} 
&=& \frac{\frac{\plaind}{\plaind T} \cumavsh }{-\frac{\plaind}{\plaind t}\Psurv(t)}\nonumber \\
&\simeq& 1 + 2\frac{q_2}{r}\left(1-\exp{-rt}\right)\left(1-\frac{\Psurv(T)}{\Psurv(t)}\right),
\elabel{avshape}
\end{eqnarray}
Fig.~\ref{fig_avshape_a}. Since the observable $V(t,T)$ suitably incorporates 
the condition $N(T)=0$, the result in \Eref{avshape} holds for the supercritical case as well.
At criticality, the avalanche shape is the parabola \cite{GleesonDurrett:2017,
PapanikolaouETAL:2011, BaldassarriETAL:2003,WillisPruessner:2018}
\begin{equation}
\lim_{r\to0}\avsh{t}{T} \simeq 1+2\frac{(q_2T)^2}{1+q_2T}\left(1-\frac{t}{T}\right)\frac{t}{T}.
\elabel{avshape0}
\end{equation}
The avalanche shape $\avsh{t}{T}$ in \Eref{avshape} is a symmetric
function with its maximum at $t=T/2$, which is bounded
\cite{BaldassarriETAL:2003} by
\begin{equation}
\lim_{T\to\infty}\avsh{\frac{T}{2}}{T} \simeq 1+ 2\frac{q_2}{r}.
\elabel{avshape_bound}
\end{equation}

\subsection{Connected correlation function $\cov{N(t_1)}{N(t_2)}$ \label{sec:twotimecorr}}

\begin{figure}
\includegraphics[width=\columnwidth]{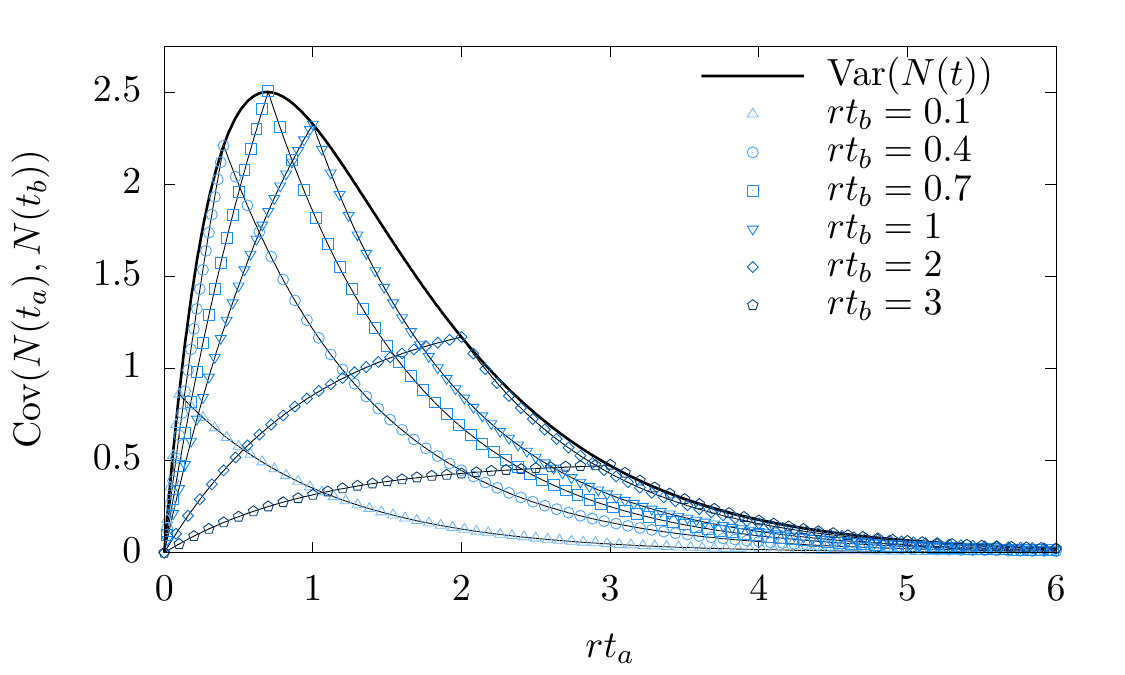}
\caption{\label{fig_corr} Two-point correlation function 
$\cov{N(t_a)}{N(t_b)}$ of the binary continuous-time branching process
with $r=10^{-1}$ and $s=1$. Our numerical results shown as symbols are in perfect agreement with the
exact expression in \Eref{2pointcf} with $t_1=\text{min}(t_a,t_b)$ and 
$t_2=\text{max}(t_a,t_b)$ (solid lines). We also show 
$\text{Var}(N(t))=\cov{N(t)}{N(t)}$, which is the envelope.}
\end{figure}

To calculate the 
expectation $\ave{ N(t_1)N(t_2)}$ we assume $0<t_1<t_2$
without loss of generality,
\begin{subeqnarray}{}
\ave{ N(t_1)N(t_2)}&\!=\!& \sandwich{\sun}{\creat\annih e^{-\hat{\mA}(t_2-\tmin)}\creat\annih e^{-\hat{\mA} \tmin}\creat}{0}\\
&=\!&\ave{\phi(t_2)\phi^\dagger(\tmin)\phi(\tmin)\phi^\dagger(0)}\\
&=\!&\ave{\phi(t_2)\tildephi(\tmin)\phi(\tmin)\tildephi(0)}
\!+\!\ave{\phi(t_2)\phi(\tmin)\tildephi(0)}\nonumber\\
&\corresponding&
\tikz[baseline=-2.5pt]{
\draw[Aactivity] (-1,0) -- (-0.1,0) node[at end,above] { }; 
\draw[Aactivity] (0.1,0) -- (1,0) node[at end,above] { }; 
}
+ 2
\tikz[baseline=-2.5pt]{
\draw[Aactivity] (0.5,0) -- (0,0) node[at end,above] { };
\draw[Aactivity] (130:0.5) -- (0,0);
\draw[Aactivity] (-130:0.5) -- (0,0);}.
\elabel{second_moment_N}
\end{subeqnarray}
The diagram on the left consists of two separate components. We refer to diagrams of that kind as disconnected diagrams, in contrast to connected diagrams that only consist of one component as the one appearing on the right. The connected correlation function is
\begin{eqnarray}{}
&&\cov{N(t_1)}{N(t_2)} 
=\ave{ N(t_1)N(t_2)} - \ave{ N(t_1)}\ave{N(t_2)} \nonumber\\
 &&\quad = \left(2\frac{q_2}{r}+1\right)\exp{-r(t_1+t_2)}
 \left(\exp{r \tmin}-1\right) 
\elabel{2pointcf}
\end{eqnarray}
which is an exact result independent of the type of branching process,
(\ie irrespective of the offspring distribution),
Fig.~\ref{fig_corr}. In particular, the variance is 
$\text{Var}\left(N(t)\right)=\cov{N(t)}{N(t)}$ \cite{Taeuber:2014}.

\subsection{\label{sec:n_pnt_corr} $n$-point correlation function}
We call $\zeta_n(t_1,\dots, t_n)$, with $0<t_1,\dots, t_n$ (not necessarily in
order),  the contribution of all binary, and therefore connected,
 diagrams to the $n$-point correlation function, where the error term is controlled as
\begin{equation}
\ave{N(t_1)\ldots N(t_n)} = \zeta_n(t_1,\dots,t_n)
+ \mO\left(\left(1-{\langle \kappa \rangle}\right)^{-(n-2)}\right).
\elabel{approxncorr}
\end{equation}
The leading order contribution $\zeta_n$ satisfies the following recurrence relation,
\begin{subeqnarray}{}
&&\zeta_n(t_1,\dots,t_n)\elabel{npointcf}\\
&&= 
\sum_{m=1}^{n-1}
\sum\limits_{\substack{\sigma\subset\{t_1,\dots,t_n\}\\|\sigma|=m}}
\left\{
\tikz[baseline=-2.5pt]{
\draw[Aactivity] (0.75,0) -- (0,0);
\begin{scope}
  \node (b) at (150:1.5) {};
  \path [postaction={decorate,decoration={raise=0ex,text along path, text align={center}, text={|\large|....}}}] (b)+(170:0.7cm) arc (170:210:0.8cm);
  \draw[Aactivity] (b)+(-150:0.2) -- ++(-150:0.8);
  \draw[Aactivity] (b)+(170:0.2) -- ++(170:0.8);
  \draw[Aactivity] (b)+(150:0.2) -- ++(150:0.8);
\end{scope}
\draw[Aactivity] (b)+(0:0.2) -- (0,0);
\draw[thick,fill=white] (b)+(0,0) circle (0.2cm);
\node at (160:2.75) {$\sigma$};
\begin{scope}
  \node (c) at (-150:1.5) {};
  \path [postaction={decorate,decoration={raise=0ex,text along path, text align={center}, text={|\large|....}}}] (c)+(170:0.7cm) arc (170:210:0.8cm);
  \draw[Aactivity] (c)+(-150:0.2) -- ++(-150:0.8);
  \draw[Aactivity] (c)+(170:0.2) -- ++(170:0.8);
  \draw[Aactivity] (c)+(150:0.2) -- ++(150:0.8);
\end{scope}
\draw[Aactivity] (c)+(0:0.2) -- (0,0);
\draw[thick,fill=white] (c)+(0,0) circle (0.2cm);
\node at (-160:2.75) {$\sigma^\text{C}$};
}
\right\}\\
&&=q_2\sum_{m=1}^{n-1}
\sum\limits_{\substack{\sigma\subset\{t_1,\dots,t_n\}\\|\sigma|=m}}
\int\limits_{0}^{\xtmin}
\zeta_{m}\bigl(t_{\sigma(1)}-t',\dots,t_{\sigma(m)}-t'\bigr)\nonumber\\
&&\quad \times\zeta_{n-m}\bigl(t_{\sigma^c(m+1)}-t',\dots,t_{\sigma^c(n)}-t'\bigr)e^{-rt'}\plaind t', \elabel{ncorr}
\end{subeqnarray}
with $\zeta_0=0$ and 
$\zeta_1(t)=e^{-rt}$, and $\xtmin=\text{min}\{t_1,\dots,t_n\}$.
Here, $\sigma$ is a subset of the set of times $\{t_1,\dots, t_n\}$, whose
size is $|\sigma|$, and $\sigma(1),\dots,\sigma(m)$ is a list of its distinct 
elements. Its complementary set is $\sigma^c=\{t_1,\dots,t_n\}\backslash\sigma$,
which contains the elements $\sigma^c(m+1),\dots,\sigma^c(n)$. \Eref{ncorr}
is symmetric under exchange of any permutation of the times $t_1,\dots, t_n$,
see the $3$-point correlation function in \Aref{3corr}.
  
This approximation is two-fold. First, it neglects higher order branching vertices proportional to $q_{j \geq 3}$, and
 secondly, it neglects contributions from disconnected diagrams, cf.~\Eref{second_moment_N}.
  Latter contributions are dominant only when $t_\text{max}=\text{max}\{t_1,\dots,t_n\} $ is smaller than 
  $s^{-1}$. For times in $\left(0,s^{-1}\right)$, the branching process
  has typically not yet undergone a change in the particle number.

\subsection{Distribution of the total avalanche size $S$}

\begin{figure}
\includegraphics[width=\columnwidth]{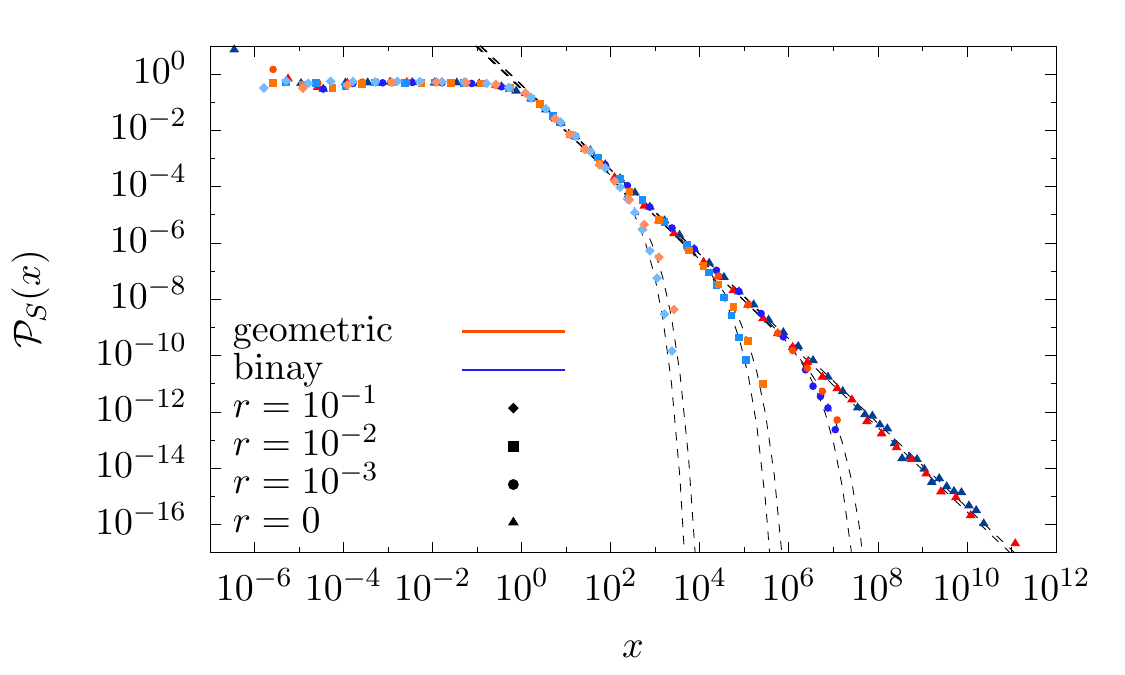}
\caption{Probability density function of the total avalanche size
$\pdf{S}{x}$ for the binary branching process (blue) and the branching process
with geometric distribution of offspring (orange), with
$r\in\left\{0,10^{-3},10^{-2},10^{-1}\right\}$ and $s=1$. Dashed lines indicate our
approximation in \Eref{pdfS}. This approximation is not valid for small
times, which explains the disagreement between the numerical results and
the dashed lines for small values of $x$.
\label{Fig:pdfS}}
\end{figure}

We define the total avalanche size as the time-integrated activity
$S = s\int\plaind t \Num(t) $.
Using $\ave{N(t)}=e^{-rt}$ and \Eref{2pointcf}, the first 
and second moments of the total avalanche size 
\cite{AschwandenCorralFontClosCh5:2013,CorralETAL:2018} 
read
\begin{subeqnarray}{}
\ave{S}&=&s\!\!\int\!\!\plaind t \ave{N(t)} =\frac{s}{r}=\frac{1}{1-{\langle \kappa \rangle}},\\
\ave{ S^2} &=&s^2\!\!\int\!\!\plaind t_1\plaind t_2  \ave{ N(t_1)N(t_2)} \!=\!\frac{s^2}{r^2}\!\left(\frac{q_2}{r}+1\right).
\elabel{Smoms2}
\end{subeqnarray}
To calculate $\ave{S^n}$ close to criticality, we use the 
approximation to the $n$-point correlation function defined in \Eref{approxncorr} and find the following recurrence relation,
\begin{subeqnarray}{}
\ave{S^n} &\simeq& s^n\int\plaind t_1\dots \plaind t_n\,\zeta_n(t_1,\dots,t_n)\\
&\simeq&\frac{q_2}{r}\sum\limits_{m=1}^{n-1}{n\choose m}\ave{S^m}\ave{S^{n-m}}\\
&\simeq&\frac{s^nq_2^{n-1}}{r^{2n-1}}2^{n-1}(2n-3)!!,\elabel{Size-Moments}
\end{subeqnarray} 
see \Aref{induction_proof} for a proof by induction of \Eref{Size-Moments}.
Similarly to \Eref{TimeMomentRatio}, we find the universal 
constant ratios of the moments of $S$,
\begin{equation}
\frac{\ave{S^k}\ave{S^\ell}}{\ave{S^{k-m}}\ave{S^{\ell+m}}}
=\frac{(2k-3)!!(2\ell-3)!!}{(2(k-m)-3)!!(2(\ell+m)-3)!!}
\end{equation}
with $k,\ell\in\mathbb{N}$ and $m\in\{0,\dots,k-1\}$.
The moment generating function of $S$ is
\begin{equation}
\mathcal{M}_S(z)\simeq 1+\frac{r-\sqrt{r^2-4sq_2z}}{2q_2},
\elabel{mgfS}
\end{equation}
and its probability density function $\pdf{S}{x}$ is the inverse
Laplace transform of $\mathcal{M}_S(-z)$,
\begin{equation}
\pdf{S}{x} \simeq \frac{1}{2}\sqrt{\frac{s}{q_2\pi}}x^{-\frac{3}{2}}\exp{-\frac{r^2x}{4q_2s}},
\elabel{pdfS}
\end{equation}
which is a power law with exponent $-3/2$ with exponential decay,
Fig.~\ref{Fig:pdfS}. At criticality, this distribution is a pure
power law.

The approximation used to derive these results, \Eref{approxncorr}, consists in neglecting contributions of disconnected
 diagrams to the $n$-point correlation function. This approximation is unjustified for total avalanche sizes corresponding
 those realisations of branching processes that underwent no branching but a single extinction event, 
 and whose sizes are therefore typically smaller than 1, because their $n$-point correlation functions
  $\ave{N(t_1)\ldots N(t_n)}$ vanish for $t_{\mathrm{max}} \gtrsim s^{-1}$. Consequently, the $n$-point correlation 
  functions are dominated by purely disconnected 
 diagrams (cf.~Sec.~\ref{sec:n_pnt_corr}). We therefore expect a breakdown of our approximation around $x = 1$. All 
 three features of the distribution of the total avalanche size, the power-law behaviour, the exponential cutoff, and the 
 breakdown of the approximation for $x < 1$, are in good agreement with numerical simulations as shown in 
 Fig.~\ref{Fig:pdfS}.


\section{Discussion and conclusions}
\label{sec:6DC}
In this paper we study the continuous-time branching process following a 
field-theoretic approach. We build on the wealth of existing results in the 
literature obtained 
through other methods. Here, we demonstrate that the Doi-Peliti field theory 
provides an elegant, intuitive, and seemingly natural language for continuous-time 
branching processes.  

We illustrate how to use the field theory to calculate a number of relevant observables,
listed in Table \ref{tab1}. Our results are valid for any offspring distribution
in the vicinity of the critical point and at large times. However, many of the
results are exact for the binary branching process and others are exact for
any branching process.
In principle, many observables can be calculated systematically using the
field theory for any offspring distribution, for any time and any
parameter set. 

In this paper, we extend the existing results in the literature
by finding explicit scaling functions and universal moment ratios for any offspring distribution. We find that all the scaling laws derived above depend on two parameters only, namely $r$ and $q_2$. Therefore, one may argue that the master equation of any branching process close to the critical point and asymptotically in large times
is captured by the action \Eref{action} with couplings $r$ and $q_2$ only.

Having established the field-theoretic ground work, in particular the basic
formalism and range of relevant observables, we may now proceed by 
extending the basic branching process into more sophisticated models
of natural phenomena. We hope that the methods established in this paper will
help reaching new boughs, branches, and twigs of the many offspring
of branching processes.

\appendix
\section{Exact expressions\label{sec:exactexpressions}}
The continuous-time branching process is exactly solvable, that is, in principle, 
all moments and correlation functions can be 
calculated in exact form if all the terms in the (possibly infinite) sums
are taken into account. Here we show some exact expressions.
The exact first three moments of $N(t)$ are
\begin{subeqnarray}{ }
\ave{\Num(t)}  &=& \exp{-rt}, \elabel{Nmom1} \\
\ave{\Num^2(t)}  &=& {\exp{-rt}} \left( 1+\frac{2q_2}{r}\left(1 - \exp{-rt}\right)\right),\\
\ave{\Num^3(t)}&=&e^{-3rt}\Bigl(\frac{6q_2^2}{r^2}-\frac{3q_3}{r}\Bigr)-e^{-2rt}\Bigl(\frac{12q_2^2}{r^2}+\frac{6q_2}{r}\Bigr)+\nonumber\\
&&+e^{-rt}\Bigl(\frac{6q_2^2}{r^2}+\frac{3q_3}{r}+\frac{6q_2}{r}+1\Bigr),
\end{subeqnarray}
and therefore the variance is
\begin{equation}
\var{N(t)} = \left(1+2\frac{q_2}{r}\right) \exp{-rt} \left(1- \exp{-rt} \right),
\elabel{varnt}
\end{equation}
which is consistent with \Eref{2pointcf} and 
\cite{DickmanVidigal:2003,Taeuber:2014,WeiPruessner:2016,Harris:1963,pazsit2007neutron}.
\label{sec:3corr}
The three-point correlation
function is, assuming $0\leq t_1\leq t_2\leq t_3$ and using \Eref{ncorr},
\begin{eqnarray}
&&\ave{N(t_1)N(t_2)N(t_3)}\simeq\zeta(t_1,t_2,t_3)\nonumber\\
&&=2\left(\frac{q_2}{r}\right)^2e^{-r(t_1+t_2+t_3)}\elabel{3ptcorr}\\
&&\quad\times\left(\left(e^{rt_1}-1\right)
\left(2e^{rt_1}+e^{rt_2}\right)-\frac{3}{2}\left(e^{2rt_1}-1\right)\right).\nonumber
\end{eqnarray}

\section{Diagrammatic representation and closed form expressions of \Esref{IIn} and \eref{In}
\label{sec:diagrams}}
Defining
\begin{equation}
a=\frac{e^{-r(T-t)}-e^{-rT}}{1-e^{-r(T-t)}},
\end{equation}
we have, firstly \eref{IIn},
\begin{subeqnarray}{}
&&\ave{\phi^n(T)\phi(t)\tildephi(0)}=
\sum_{k=1}^n\sum_{m_1,\ldots,m_k}\binom{n}{m_1,\ldots,m_k}\nonumber\\
&&\qquad\times
g_{m_1}(T-t)\cdots g_{m_k}(T-t)g_{k+1}(t)\frac{1}{k!}\\
 &&\corresponding
\sum_{k=1}^n\sum_{m_1,\ldots,m_k}\binom{n}{m_1,\ldots,m_k}
\left\{
\tikz[baseline=-2.5pt,scale=0.9]{
\draw[thick,fill=white] (0,0) circle (0.2cm);
\draw[Aactivity] (0.75,0) -- (0.2,0);
  \path [postaction={decorate,decoration={raise=0ex,text along path, text align={center}, text={|\large|......}}}] (145:0.7cm) arc (145:170:0.8cm);
  \node at (155:0.9) {$k$};
  \node (a) at (175:1.5) {};
  \path [postaction={decorate,decoration={raise=0ex,text along path, text align={center}, text={|\large|....}}}] (a)+(170:0.7cm) arc (170:210:0.8cm);
  \draw[Aactivity] (a)+(-150:0.2) -- ++(-150:0.8);
  \draw[Aactivity] (a)+(170:0.2) -- ++(170:0.8);
  \draw[Aactivity] (a)+(150:0.2) -- ++(150:0.8);
\draw[Aactivity] (a)+(0:0.2) -- (175:0.2);
\draw[thick,fill=white] (a)+(0,0) circle (0.2cm);
\node at (180:2.5) {$m_k$};
\begin{scope}
  \node (b) at (140:1.75) {};
  \path [postaction={decorate,decoration={raise=0ex,text along path, text align={center}, text={|\large|....}}}] (b)+(170:0.7cm) arc (170:210:0.8cm);
  \draw[Aactivity] (b)+(-150:0.2) -- ++(-150:0.8);
  \draw[Aactivity] (b)+(170:0.2) -- ++(170:0.8);
  \draw[Aactivity] (b)+(150:0.2) -- ++(150:0.8);
\end{scope}
\draw[Aactivity] (b)+(0:0.2) -- (140:0.2);
\draw[thick,fill=white] (b)+(0,0) circle (0.2cm);
\node at (160:2.5) {$m_1$};
\begin{scope}
  \node (c) at (-150:1.1) {};
\end{scope}
\draw[Aactivity] (c)+(0:0.2) -- (-150:0.2);
}\right\}
\nonumber\\
&&=n!e^{-rt}\left(\frac{q_2}{r}\right)^n\!\!
\left(1-e^{-rT}\right)^n\!
\left[\frac{a^2(n-1)}{(1+a)^2}+\frac{2a}{1+a}\right],
\end{subeqnarray}
and secondly \eref{In},
\begin{subeqnarray}{}
&&\ave{\phi^n(T)\tildephi(t)\phi(t)\tildephi(0)}
=\sum_{k=1}^n\sum_{m_1,\ldots,m_k}\binom{n}{m_1,\ldots,m_k} \nonumber\\
&&\qquad\times g_{m_1}(T-t)\cdots g_{m_k}(T-t)g_{k}(t)\frac{1}{(k-1)!} \\
&&\corresponding
\sum_{k=1}^n\sum_{m_1,\ldots,m_k}\binom{n}{m_1,\ldots,m_k}
\left\{
\tikz[baseline=-2.5pt,scale=0.9]{
\draw[thick,fill=white] (0,0) circle (0.2cm);
\draw[Aactivity] (0.75,0) -- (0.2,0);
  \path [postaction={decorate,decoration={raise=0ex,text along path, text align={center}, text={|\large|......}}}] (170:0.7cm) arc (170:220:0.8cm);
  \node at (190:0.9) {$k$};
  \node (a) at (170:1.5) {};
  \path [postaction={decorate,decoration={raise=0ex,text along path, text align={center}, text={|\large|....}}}] (a)+(170:0.7cm) arc (170:210:0.8cm);
  \draw[Aactivity] (a)+(-150:0.2) -- ++(-150:0.8);
  \draw[Aactivity] (a)+(170:0.2) -- ++(170:0.8);
  \draw[Aactivity] (a)+(150:0.2) -- ++(150:0.8);
\draw[Aactivity] (a)+(0:0.2) -- (170:0.2);
\draw[thick,fill=white] (a)+(0,0) circle (0.2cm);
\node at (180:2.5) {$m_2$};
\begin{scope}
  \node (b) at (140:1.75) {};
  \path [postaction={decorate,decoration={raise=0ex,text along path, text align={center}, text={|\large|....}}}] (b)+(170:0.7cm) arc (170:210:0.8cm);
  \draw[Aactivity] (b)+(-150:0.2) -- ++(-150:0.8);
  \draw[Aactivity] (b)+(170:0.2) -- ++(170:0.8);
  \draw[Aactivity] (b)+(150:0.2) -- ++(150:0.8);
\end{scope}
\draw[Aactivity] (b)+(0:0.2) -- (140:0.2);
\draw[thick,fill=white] (b)+(0,0) circle (0.2cm);
\node at (160:2.5) {$m_1$};
\begin{scope}
  \node (c) at (-140:1.75) {};
  \path [postaction={decorate,decoration={raise=0ex,text along path, text align={center}, text={|\large|....}}}] (c)+(170:0.7cm) arc (170:210:0.8cm);
  \draw[Aactivity] (c)+(-150:0.2) -- ++(-150:0.8);
  \draw[Aactivity] (c)+(170:0.2) -- ++(170:0.8);
  \draw[Aactivity] (c)+(150:0.2) -- ++(150:0.8);
\end{scope}
\draw[Aactivity] (-140:1.75) -- (-140:1.05);
\draw[Aactivity] (-140:0.85) -- (-140:0.2);
\draw[thick,fill=white] (c)+(0,0) circle (0.2cm);
\node at (207:2.6) {$m_k$};
}\right\}\nonumber\\
&&=\frac{n!e^{-rt}}{1-e^{-rt}}\left(\frac{q_2}{r}\right)^{n-1}
\left(1-e^{-rT}\right)^n
\nonumber\\
&&\qquad\times\left[\frac{a^2(n-1)}{(1+a)^2}+\frac{a}{1+a}\right].
\end{subeqnarray}

\section{Averaged avalanche shape}
\label{sec:average_shape}
In \Sref{avsh}, we derive analytically the expected avalanche shape $V(t,T)$ for a specific time of death $T$. However, direct comparison with numerics is computationally very expensive as specific large times of death occur rarely for subcritical branching processes.
Here we describe an observable that is accessible both analytically and
numerically: the averaged avalanche shape $\ave{V(\tau)}_T$. For a fixed parameter
set, we first rescale time $\tau=t/T$ 
and then average all the avalanche profiles irrespectively of $T$. Finally,
to achieve convergence to a shape comparable across parameter
settings, we normalise the result by area \cite{WillisPruessner:2018},
\begin{subeqnarray}
\ave{V(\tau)}_T&=&\frac{1}{N_V}\int \limits_0^{\infty}\plaind T\mathcal{P}_T(T)V(\tau T,T),\\
N_V&=&\int\limits_0^1\int\limits_0^{\infty}\plaind \tau\plaind T\mathcal{P}_T(T)V(\tau T,T).
\end{subeqnarray}
The result \cite{Mathematica:11.3.0.0} 
can be expressed with the Gaussian hypergeometric function ${_2F}_1(a,b,c,z)$,
\begin{equation}
\ave{V(\tau)}_T=\frac{1}{N_V}+\tau(\tau-1)\frac{q_2\,F(\tau,q_2,r)}{(q_2+r)N_V},\elabel{average_avshape}
\end{equation}
where
\begin{eqnarray}
F(\tau,q_2,r)&=&\frac{{_2F}_1\left(1,2-\tau,3-\tau,\frac{q_2}{q_2+r}\right)}{\tau-2}\nonumber\\
&&-\frac{{_2F}_1\left(1,1+\tau,2+\tau,\frac{q_2}{q_2+r}\right)}{\tau+1}.
\end{eqnarray}
Both $F$ and $N_V$ diverge at the critical point with the limit 
\begin{align}
\lim\limits_{r\rightarrow0}\frac{F(\tau,q_2,r)}{N_V}=6
\end{align}

\section{Proof of \Eref{Size-Moments}\label{sec:induction_proof}}
\Eref{Size-Moments} can be proved by induction. In \Eref{Smoms2} we see
that it applies to $\ave{S}$. The approximation of binary tree diagrams of 
$\ave{N(t_1)N(t_2)}$ gives $\ave{S^2}=s^2q_2/r^3$, which also satisfies 
\Eref{Size-Moments}. 
The induction step is verified by
\begin{eqnarray}{}\elabel{recurrenceInduction}
\ave{S^n}&=&\frac{q_2}{r}\sum\limits_{m=1}^{n-1}\binom{n}{m}\left(\frac{s^mq_2^{m-1}2^{m-1}(2m-3)!!}{r^{2m-1}}\right)\\
&&\qquad\times\left(\frac{s^{n-m}q_2^{n-m-1}2^{n-m-1}(2(n-m)-3)!!}{r^{2(n-m)-1}}\right)\nonumber\\
&=&\frac{s^nq_2^{n-1}}{r^{2n-1}}2^{n-2}\sum\limits_{m=1}^{n-1}\binom{n}{m}(2m-3)!!(2(n-m)-3)!! .\nonumber
\end{eqnarray}
This sum is equivalent to
\begin{subeqnarray}{}
&&\sum\limits_{m=1}^{n-1}\binom{n}{m}(2m-3)!!(2(n-m)-3)!!\\
&&= \frac{1}{n-1}\sum\limits_{m=1}^{n-1}\binom{n}{m}(2m-3)!!(2(n-m)-1)!!\nonumber\\
&&=\frac{1}{n-1}\sum\limits_{k=0}^{n-2}\binom{n}{k+1}(2k-1)!!(2(n-k)-3)!!\nonumber\\
&&=2(2n-3)!!,
\elabel{identityInduction}
\end{subeqnarray}
where we have used the identity \cite{callan:2009},
\begin{equation}
\sum\limits_{k=0}^{n-1}\binom{n}{k+1}(2k-1)!!(2(n-k)-3)!!=(2n-1)!!.
\end{equation}
Using \Eref{identityInduction} in \Eref{recurrenceInduction}
reproduces \Eref{Size-Moments}, thereby completing the proof.

\begin{acknowledgements}
We thank Nanxin Wei, Stephanie Miller, Kay Wiese, Ignacio Bordeu Weldt, Eric Smith, David Krakauer, and 
Nicholas Moloney, for fruitful discussions. We extend our gratitude to Andy Thomas for invaluable computing support.
\end{acknowledgements}

\bibliography{articles,books}

\begin{thebibliography}{42}%
\makeatletter
\providecommand \@ifxundefined [1]{%
 \@ifx{#1\undefined}
}%
\providecommand \@ifnum [1]{%
 \ifnum #1\expandafter \@firstoftwo
 \else \expandafter \@secondoftwo
 \fi
}%
\providecommand \@ifx [1]{%
 \ifx #1\expandafter \@firstoftwo
 \else \expandafter \@secondoftwo
 \fi
}%
\providecommand \natexlab [1]{#1}%
\providecommand \enquote  [1]{``#1''}%
\providecommand \bibnamefont  [1]{#1}%
\providecommand \bibfnamefont [1]{#1}%
\providecommand \citenamefont [1]{#1}%
\providecommand \href@noop [0]{\@secondoftwo}%
\providecommand \href [0]{\begingroup \@sanitize@url \@href}%
\providecommand \@href[1]{\@@startlink{#1}\@@href}%
\providecommand \@@href[1]{\endgroup#1\@@endlink}%
\providecommand \@sanitize@url [0]{\catcode `\\12\catcode `\$12\catcode
  `\&12\catcode `\#12\catcode `\^12\catcode `\_12\catcode `\%12\relax}%
\providecommand \@@startlink[1]{}%
\providecommand \@@endlink[0]{}%
\providecommand \url  [0]{\begingroup\@sanitize@url \@url }%
\providecommand \@url [1]{\endgroup\@href {#1}{\urlprefix }}%
\providecommand \urlprefix  [0]{URL }%
\providecommand \Eprint [0]{\href }%
\providecommand \doibase [0]{http://dx.doi.org/}%
\providecommand \selectlanguage [0]{\@gobble}%
\providecommand \bibinfo  [0]{\@secondoftwo}%
\providecommand \bibfield  [0]{\@secondoftwo}%
\providecommand \translation [1]{[#1]}%
\providecommand \BibitemOpen [0]{}%
\providecommand \bibitemStop [0]{}%
\providecommand \bibitemNoStop [0]{.\EOS\space}%
\providecommand \EOS [0]{\spacefactor3000\relax}%
\providecommand \BibitemShut  [1]{\csname bibitem#1\endcsname}%
\let\auto@bib@innerbib\@empty
\bibitem [{\citenamefont {Harris}(1963)}]{Harris:1963}%
  \BibitemOpen
  \bibfield  {author} {\bibinfo {author} {\bibfnamefont {T.~E.}\ \bibnamefont
  {Harris}},\ }\href@noop {} {\emph {\bibinfo {title} {The Theory of Branching
  Processes}}}\ (\bibinfo  {publisher} {Springer-Verlag},\ \bibinfo {address}
  {Berlin, Germany},\ \bibinfo {year} {1963})\BibitemShut {NoStop}%
\bibitem [{\citenamefont {Zapperi}\ \emph {et~al.}(1995)\citenamefont
  {Zapperi}, \citenamefont {Lauritsen},\ and\ \citenamefont
  {Stanley}}]{ZapperiLauritsenStanley:1995}%
  \BibitemOpen
  \bibfield  {author} {\bibinfo {author} {\bibfnamefont {S.}~\bibnamefont
  {Zapperi}}, \bibinfo {author} {\bibfnamefont {K.~B.}\ \bibnamefont
  {Lauritsen}}, \ and\ \bibinfo {author} {\bibfnamefont {H.~E.}\ \bibnamefont
  {Stanley}},\ }\href@noop {} {\bibfield  {journal} {\bibinfo  {journal} {Phys.
  Rev. Lett.}\ }\textbf {\bibinfo {volume} {75}},\ \bibinfo {pages} {4071}
  (\bibinfo {year} {1995})}\BibitemShut {NoStop}%
\bibitem [{\citenamefont {Lee}\ \emph {et~al.}(2004)\citenamefont {Lee},
  \citenamefont {Goh}, \citenamefont {Kahng},\ and\ \citenamefont
  {Kim}}]{lee2004branching}%
  \BibitemOpen
  \bibfield  {author} {\bibinfo {author} {\bibfnamefont {D.~S.}\ \bibnamefont
  {Lee}}, \bibinfo {author} {\bibfnamefont {K.~I.}\ \bibnamefont {Goh}},
  \bibinfo {author} {\bibfnamefont {B.}~\bibnamefont {Kahng}}, \ and\ \bibinfo
  {author} {\bibfnamefont {D.}~\bibnamefont {Kim}},\ }\href@noop {} {\bibfield
  {journal} {\bibinfo  {journal} {J. Korean Phys. Soc.}\ }\textbf {\bibinfo
  {volume} {44}},\ \bibinfo {pages} {633} (\bibinfo {year} {2004})}\BibitemShut
  {NoStop}%
\bibitem [{\citenamefont {Gleeson}\ and\ \citenamefont
  {Durrett}(2017)}]{GleesonDurrett:2017}%
  \BibitemOpen
  \bibfield  {author} {\bibinfo {author} {\bibfnamefont {J.~P.}\ \bibnamefont
  {Gleeson}}\ and\ \bibinfo {author} {\bibfnamefont {R.}~\bibnamefont
  {Durrett}},\ }\href@noop {} {\bibfield  {journal} {\bibinfo  {journal} {Nat.
  Com.}\ }\textbf {\bibinfo {volume} {8}},\ \bibinfo {pages} {1227} (\bibinfo
  {year} {2017})}\BibitemShut {NoStop}%
\bibitem [{\citenamefont {Gilbert}(1961)}]{gilbert1961random}%
  \BibitemOpen
  \bibfield  {author} {\bibinfo {author} {\bibfnamefont {E.~N.}\ \bibnamefont
  {Gilbert}},\ }\href@noop {} {\bibfield  {journal} {\bibinfo  {journal} {J.
  SIAM}\ }\textbf {\bibinfo {volume} {9}},\ \bibinfo {pages} {533} (\bibinfo
  {year} {1961})}\BibitemShut {NoStop}%
\bibitem [{\citenamefont {Durrett}(2006)}]{durrett2006random}%
  \BibitemOpen
  \bibfield  {author} {\bibinfo {author} {\bibfnamefont {R.}~\bibnamefont
  {Durrett}},\ }\href@noop {} {\enquote {\bibinfo {title} {Random graph
  dynamics, vol. 20 of cambridge series in statistical and probabilistic
  mathematics},}\ } (\bibinfo {year} {2006})\BibitemShut {NoStop}%
\bibitem [{\citenamefont {Marzocchi}\ and\ \citenamefont
  {Lombardi}(2008)}]{marzocchi2008}%
  \BibitemOpen
  \bibfield  {author} {\bibinfo {author} {\bibfnamefont {W.}~\bibnamefont
  {Marzocchi}}\ and\ \bibinfo {author} {\bibfnamefont {A.~M.}\ \bibnamefont
  {Lombardi}},\ }\href@noop {} {\bibfield  {journal} {\bibinfo  {journal}
  {J.~Geophys.~Res.}\ }\textbf {\bibinfo {volume} {113}},\ \bibinfo {pages}
  {B08317} (\bibinfo {year} {2008})}\BibitemShut {NoStop}%
\bibitem [{\citenamefont {Corral}\ and\ \citenamefont
  {Font-Clos}(2013)}]{AschwandenCorralFontClosCh5:2013}%
  \BibitemOpen
  \bibfield  {author} {\bibinfo {author} {\bibfnamefont {A.}~\bibnamefont
  {Corral}}\ and\ \bibinfo {author} {\bibfnamefont {F.}~\bibnamefont
  {Font-Clos}},\ }in\ \href@noop {} {\emph {\bibinfo {booktitle}
  {Self-Organized Criticality Systems}}},\ \bibinfo {editor} {edited by\
  \bibinfo {editor} {\bibfnamefont {M.~J.}\ \bibnamefont {Aschwanden}}}\
  (\bibinfo  {publisher} {OpenAcademicPress},\ \bibinfo {address} {Berlin,
  Germany},\ \bibinfo {year} {2013})\ pp.\ \bibinfo {pages}
  {183--228}\BibitemShut {NoStop}%
\bibitem [{\citenamefont {Reed}\ and\ \citenamefont {Hughes}(2003)}]{reed2003}%
  \BibitemOpen
  \bibfield  {author} {\bibinfo {author} {\bibfnamefont {W.~J.}\ \bibnamefont
  {Reed}}\ and\ \bibinfo {author} {\bibfnamefont {B.~D.}\ \bibnamefont
  {Hughes}},\ }\href@noop {} {\bibfield  {journal} {\bibinfo  {journal}
  {Phys.~A}\ }\textbf {\bibinfo {volume} {319}},\ \bibinfo {pages} {579}
  (\bibinfo {year} {2003})}\BibitemShut {NoStop}%
\bibitem [{\citenamefont {Kimmel}\ and\ \citenamefont
  {Axelrod}(2002)}]{kimmel2002}%
  \BibitemOpen
  \bibfield  {author} {\bibinfo {author} {\bibfnamefont {M.}~\bibnamefont
  {Kimmel}}\ and\ \bibinfo {author} {\bibfnamefont {D.~E.}\ \bibnamefont
  {Axelrod}},\ }\href@noop {} {\emph {\bibinfo {title} {Branching Processes in
  Biology (Interdisciplinary Applied Mathematics)}}},\ Vol.~\bibinfo {volume}
  {19}\ (\bibinfo  {publisher} {Springer},\ \bibinfo {address} {Berlin},\
  \bibinfo {year} {2002})\BibitemShut {NoStop}%
\bibitem [{\citenamefont {Durrett}(2015)}]{durrett2015branching}%
  \BibitemOpen
  \bibfield  {author} {\bibinfo {author} {\bibfnamefont {R.}~\bibnamefont
  {Durrett}},\ }in\ \href@noop {} {\emph {\bibinfo {booktitle} {Branching
  Process Models of Cancer}}}\ (\bibinfo  {publisher} {Springer, Berlin},\ \bibinfo
  {year} {2015}),\ pp.\ \bibinfo {pages} {1--63}\BibitemShut {NoStop}%
\bibitem [{\citenamefont {Williams}(2013)}]{williams2013random}%
  \BibitemOpen
  \bibfield  {author} {\bibinfo {author} {\bibfnamefont {M.~M.~R.}\
  \bibnamefont {Williams}},\ }\href@noop {} {\emph {\bibinfo {title} {Random
  Processes in Nuclear Reactors}}}\ (\bibinfo  {publisher} {Elsevier, Amsterdam},\
  \bibinfo {year} {2013})\BibitemShut {NoStop}%
\bibitem [{\citenamefont {P{\'a}zsit}\ and\ \citenamefont
  {P{\'a}l}(2007)}]{pazsit2007neutron}%
  \BibitemOpen
  \bibfield  {author} {\bibinfo {author} {\bibfnamefont {I.}~\bibnamefont
  {P{\'a}zsit}}\ and\ \bibinfo {author} {\bibfnamefont {L.}~\bibnamefont
  {P{\'a}l}},\ }\href@noop {} {\emph {\bibinfo {title} {Neutron Fluctuations: A
  Treatise on the Physics of Branching Processes}}}\ (\bibinfo  {publisher}
  {Elsevier, Amsterdam},\ \bibinfo {year} {2007})\BibitemShut {NoStop}%
\bibitem [{\citenamefont {Rockmore}\ \emph {et~al.}(2018)\citenamefont
  {Rockmore}, \citenamefont {Fang}, \citenamefont {Foti}, \citenamefont
  {Ginsburg},\ and\ \citenamefont {Krakauer}}]{rockmore2018}%
  \BibitemOpen
  \bibfield  {author} {\bibinfo {author} {\bibfnamefont {D.~N.}\ \bibnamefont
  {Rockmore}}, \bibinfo {author} {\bibfnamefont {C.}~\bibnamefont {Fang}},
  \bibinfo {author} {\bibfnamefont {N.~J.}\ \bibnamefont {Foti}}, \bibinfo
  {author} {\bibfnamefont {T.}~\bibnamefont {Ginsburg}}, \ and\ \bibinfo
  {author} {\bibfnamefont {D.~C.}\ \bibnamefont {Krakauer}},\ }\href@noop {}
  {\bibfield  {journal} {\bibinfo  {journal} {J.~Assoc.~Info.~Sci.~Technol.}\ }
  \textbf {\bibinfo {volume} {69}},\
  \bibinfo {pages} {483} (\bibinfo {year} {2018})}\BibitemShut {NoStop}%
\bibitem [{\citenamefont {Seshadri}\ \emph {et~al.}(2018)\citenamefont
  {Seshadri}, \citenamefont {Klaus}, \citenamefont {Winkowski}, \citenamefont
  {Kanold},\ and\ \citenamefont {Plenz}}]{seshadri2018altered}%
  \BibitemOpen
  \bibfield  {author} {\bibinfo {author} {\bibfnamefont {S.}~\bibnamefont
  {Seshadri}}, \bibinfo {author} {\bibfnamefont {A.}~\bibnamefont {Klaus}},
  \bibinfo {author} {\bibfnamefont {D.~E.}\ \bibnamefont {Winkowski}}, \bibinfo
  {author} {\bibfnamefont {P.~O.}\ \bibnamefont {Kanold}}, \ and\ \bibinfo
  {author} {\bibfnamefont {D.}~\bibnamefont {Plenz}},\ }\href@noop {}
  {\bibfield  {journal} {\bibinfo  {journal} {Translational Psychiatry}\
  }\textbf {\bibinfo {volume} {8}},\ \bibinfo {pages} {3} (\bibinfo {year}
  {2018})}\BibitemShut {NoStop}%
\bibitem [{\citenamefont {Beggs}\ and\ \citenamefont
  {Plenz}(2003)}]{BeggsPlenz:2003}%
  \BibitemOpen
  \bibfield  {author} {\bibinfo {author} {\bibfnamefont {J.~M.}\ \bibnamefont
  {Beggs}}\ and\ \bibinfo {author} {\bibfnamefont {D.}~\bibnamefont {Plenz}},\
  }\href@noop {} {\bibfield  {journal} {\bibinfo  {journal} {J. Neurosci.}\
  }\textbf {\bibinfo {volume} {23}},\ \bibinfo {pages} {11167} (\bibinfo {year}
  {2003})}\BibitemShut {NoStop}%
\bibitem [{\citenamefont {T{\"a}uber}(2014)}]{Taeuber:2014}%
  \BibitemOpen
  \bibfield  {author} {\bibinfo {author} {\bibfnamefont {U.~C.}\ \bibnamefont
  {T{\"a}uber}},\ }\href@noop {} {\emph {\bibinfo {title} {Critical
  Dynamics}}}\ (\bibinfo  {publisher} {Cambridge University Press},\ \bibinfo
  {address} {Cambridge, UK},\ \bibinfo {year} {2014})\ pp.\ \bibinfo {pages}
  {i--xvi,1--511}\BibitemShut {NoStop}%
\bibitem [{\citenamefont {Gardiner}(1997)}]{Gardiner:1997}%
  \BibitemOpen
  \bibfield  {author} {\bibinfo {author} {\bibfnamefont {C.~W.}\ \bibnamefont
  {Gardiner}},\ }\href@noop {} {\emph {\bibinfo {title} {Handbook of Stochastic
  Methods}}},\ \bibinfo {edition} {2nd}\ ed.\ (\bibinfo  {publisher}
  {Springer-Verlag},\ \bibinfo {address} {Berlin, Germany},\ \bibinfo {year}
  {1997})\BibitemShut {NoStop}%
\bibitem [{\citenamefont {Grimmett}\ and\ \citenamefont
  {Stirzaker}(1992)}]{GrimmettStirzaker:1992}%
  \BibitemOpen
  \bibfield  {author} {\bibinfo {author} {\bibfnamefont {G.~R.}\ \bibnamefont
  {Grimmett}}\ and\ \bibinfo {author} {\bibfnamefont {D.~R.}\ \bibnamefont
  {Stirzaker}},\ }\href@noop {} {\emph {\bibinfo {title} {Probability and
  Random Processes}}},\ \bibinfo {edition} {2nd}\ ed.\ (\bibinfo  {publisher}
  {Oxford University Press},\ \bibinfo {address} {New York, NY, USA},\ \bibinfo
  {year} {1992})\BibitemShut {NoStop}%
\bibitem [{\citenamefont {Pal}(1958)}]{pal1958theory}%
  \BibitemOpen
  \bibfield  {author} {\bibinfo {author} {\bibfnamefont {L.}~\bibnamefont
  {Pal}},\ }\href@noop {} {\bibfield  {journal} {\bibinfo  {journal} {Nuovo
  Cim.}\ }\textbf {\bibinfo {volume} {7}},\ \bibinfo {pages} {25} (\bibinfo
  {year} {1958})}\BibitemShut {NoStop}%
\bibitem [{\citenamefont {Bell}(1965)}]{bell1965stochastic}%
  \BibitemOpen
  \bibfield  {author} {\bibinfo {author} {\bibfnamefont {G.~I.}\ \bibnamefont
  {Bell}},\ }\href@noop {} {\bibfield  {journal} {\bibinfo  {journal}
  {Nucl.~Sci.~and Engr.~}\ }\textbf {\bibinfo {volume} {21}},\ \bibinfo {pages}
  {390} (\bibinfo {year} {1965})}\BibitemShut {NoStop}%
\bibitem [{\citenamefont {Kitamura}\ and\ \citenamefont
  {Misawa}(2019)}]{kitamura2019delayed}%
  \BibitemOpen
  \bibfield  {author} {\bibinfo {author} {\bibfnamefont {Y.}~\bibnamefont
  {Kitamura}}\ and\ \bibinfo {author} {\bibfnamefont {T.}~\bibnamefont
  {Misawa}},\ }\href@noop {} {\bibfield  {journal} {\bibinfo  {journal} {Ann.
  Nucl. Energy}\ }\textbf {\bibinfo {volume} {123}},\ \bibinfo {pages} {119}
  (\bibinfo {year} {2019})}\BibitemShut {NoStop}%
\bibitem [{\citenamefont {Le~Gall}(2005)}]{LeGall:2005}%
  \BibitemOpen
  \bibfield  {author} {\bibinfo {author} {\bibfnamefont {J.-F.}\ \bibnamefont
  {Le~Gall}},\ }\href@noop {} {\bibfield  {journal} {\bibinfo  {journal}
  {Probab.~Surv.}\ }\textbf {\bibinfo {volume} {2}},\ \bibinfo {pages} {245}
  (\bibinfo {year} {2005})}\BibitemShut {NoStop}%
\bibitem [{\citenamefont {Aldous}(1993)}]{Aldous:1993}%
  \BibitemOpen
  \bibfield  {author} {\bibinfo {author} {\bibfnamefont {D.}~\bibnamefont
  {Aldous}},\ }\href@noop {} {\bibfield  {journal} {\bibinfo  {journal}
  {Ann.~Probab.}\ }\textbf {\bibinfo {volume} {21}},\ \bibinfo {pages} {248}
  (\bibinfo {year} {1993})}\BibitemShut {NoStop}%
\bibitem [{\citenamefont {Doi}(1976)}]{Doi:1976}%
  \BibitemOpen
  \bibfield  {author} {\bibinfo {author} {\bibfnamefont {M.}~\bibnamefont
  {Doi}},\ }\href@noop {} {\bibfield  {journal} {\bibinfo  {journal} {J. Phys.
  A: Math. Gen.}\ }\textbf {\bibinfo {volume} {9}},\ \bibinfo {pages} {1465}
  (\bibinfo {year} {1976})}\BibitemShut {NoStop}%
\bibitem [{\citenamefont {Peliti}(1985)}]{Peliti:1985}%
  \BibitemOpen
  \bibfield  {author} {\bibinfo {author} {\bibfnamefont {L.}~\bibnamefont
  {Peliti}},\ }\href@noop {} {\bibfield  {journal} {\bibinfo  {journal} {J.
  Phys. (Paris)}\ }\textbf {\bibinfo {volume} {46}},\ \bibinfo {pages} {1469}
  (\bibinfo {year} {1985})}\BibitemShut {NoStop}%
\bibitem [{\citenamefont {Dickman}\ and\ \citenamefont
  {Vidigal}(2003)}]{DickmanVidigal:2003}%
  \BibitemOpen
  \bibfield  {author} {\bibinfo {author} {\bibfnamefont {R.}~\bibnamefont
  {Dickman}}\ and\ \bibinfo {author} {\bibfnamefont {R.}~\bibnamefont
  {Vidigal}},\ }\href@noop {} {\bibfield  {journal} {\bibinfo  {journal} {Braz.
  J. Phys.}\ }\textbf {\bibinfo {volume} {33}},\ \bibinfo {pages} {73}
  (\bibinfo {year} {2003})}\BibitemShut {NoStop}%
\bibitem [{\citenamefont {T{\"a}uber}\ \emph {et~al.}(2005)\citenamefont
  {T{\"a}uber}, \citenamefont {Howard},\ and\ \citenamefont
  {Vollmayr-Lee}}]{TaeuberHowardVollmayr-Lee:2005}%
  \BibitemOpen
  \bibfield  {author} {\bibinfo {author} {\bibfnamefont {U.~C.}\ \bibnamefont
  {T{\"a}uber}}, \bibinfo {author} {\bibfnamefont {M.}~\bibnamefont {Howard}},
  \ and\ \bibinfo {author} {\bibfnamefont {B.~P.}\ \bibnamefont
  {Vollmayr-Lee}},\ }\href@noop {} {\bibfield  {journal} {\bibinfo  {journal}
  {J. Phys. A: Math. Gen.}\ }\textbf {\bibinfo {volume} {38}},\ \bibinfo
  {pages} {R79} (\bibinfo {year} {2005})}\BibitemShut {NoStop}%
\bibitem [{Note1()}]{Note1}%
  \BibitemOpen
  \bibinfo {note} {The Stirling numbers of the second kind can be calculated
  using the expression \begin {equation*} \begin {Bmatrix}{n}\\{\ell }\end
  {Bmatrix} = \protect \frac {1}{\ell !}\DOTSB \sum@ \slimits@ _{j=0}^\ell
  (-1)^{\ell -j}\protect \binom {\ell }{j}j^n. \end {equation*}}\BibitemShut
  {NoStop}%
\bibitem [{Note2()}]{Note2}%
  \BibitemOpen
  \bibinfo {note} {The Stirling numbers of the second kind satisfy the identity
  $$ N^n = \DOTSB \sum@ \slimits@ \limits _{\ell =0}^{n}\begin
  {Bmatrix}{n}\\{\ell }\end {Bmatrix} N(N-1)\protect \ldots (N-\ell +1) .
  $$}\BibitemShut {NoStop}%
\bibitem [{\citenamefont {Garcia-Millan}\ \emph {et~al.}(2015)\citenamefont
  {Garcia-Millan}, \citenamefont {Font-Clos},\ and\ \citenamefont
  {Corral}}]{Garcia-MillanFont-ClosCorral:2016}%
  \BibitemOpen
  \bibfield  {author} {\bibinfo {author} {\bibfnamefont {R.}~\bibnamefont
  {Garcia-Millan}}, \bibinfo {author} {\bibfnamefont {F.}~\bibnamefont
  {Font-Clos}}, \ and\ \bibinfo {author} {\bibfnamefont {A.}~\bibnamefont
  {Corral}},\ }\href@noop {} {\bibfield  {journal} {\bibinfo  {journal} {Phys.
  Rev. E}\ }\textbf {\bibinfo {volume} {91}},\ \bibinfo {pages} {042122}
  (\bibinfo {year} {2015})}\BibitemShut {NoStop}%
\bibitem [{\citenamefont {Wei}\ and\ \citenamefont
  {Pruessner}(2016)}]{WeiPruessner:2016}%
  \BibitemOpen
  \bibfield  {author} {\bibinfo {author} {\bibfnamefont {N.}~\bibnamefont
  {Wei}}\ and\ \bibinfo {author} {\bibfnamefont {G.}~\bibnamefont
  {Pruessner}},\ }\href@noop {} {\bibfield  {journal} {\bibinfo  {journal}
  {Phys. Rev. E}\ }\textbf {\bibinfo {volume} {94}},\ \bibinfo {pages} {066101}
  (\bibinfo {year} {2016})},\ \bibinfo {note} {comment on
  \citep{Garcia-MillanFont-ClosCorral:2016}}\BibitemShut {NoStop}%
\bibitem [{\citenamefont {Corral}\ \emph {et~al.}(2016)\citenamefont {Corral},
  \citenamefont {Garcia-Millan},\ and\ \citenamefont
  {Font-Clos}}]{corral2016exact}%
  \BibitemOpen
  \bibfield  {author} {\bibinfo {author} {\bibfnamefont {{\'A}.}~\bibnamefont
  {Corral}}, \bibinfo {author} {\bibfnamefont {R.}~\bibnamefont
  {Garcia-Millan}}, \ and\ \bibinfo {author} {\bibfnamefont {F.}~\bibnamefont
  {Font-Clos}},\ }\href@noop {} {\bibfield  {journal} {\bibinfo  {journal}
  {PloS One}\ }\textbf {\bibinfo {volume} {11}},\ \bibinfo {pages} {e0161586}
  (\bibinfo {year} {2016})}\BibitemShut {NoStop}%
\bibitem [{\citenamefont {Dobrinevski}\ \emph {et~al.}(2015)\citenamefont
  {Dobrinevski}, \citenamefont {Le~Doussal},\ and\ \citenamefont
  {Wiese}}]{DobrinevskiETAL:2015}%
  \BibitemOpen
  \bibfield  {author} {\bibinfo {author} {\bibfnamefont {A.}~\bibnamefont
  {Dobrinevski}}, \bibinfo {author} {\bibfnamefont {P.}~\bibnamefont
  {Le~Doussal}}, \ and\ \bibinfo {author} {\bibfnamefont {K.~J.}\ \bibnamefont
  {Wiese}},\ }\href@noop {} {\bibfield  {journal} {\bibinfo  {journal}
  {Europhys. Lett.}\ }\textbf {\bibinfo {volume} {108}},\ \bibinfo {pages}
  {66002} (\bibinfo {year} {2015})}\BibitemShut {NoStop}%
\bibitem [{\citenamefont {Sheikh}\ \emph {et~al.}(2016)\citenamefont {Sheikh},
  \citenamefont {Weaver},\ and\ \citenamefont {Dahmen}}]{SheikhETAL:2016}%
  \BibitemOpen
  \bibfield  {author} {\bibinfo {author} {\bibfnamefont {M.~A.}\ \bibnamefont
  {Sheikh}}, \bibinfo {author} {\bibfnamefont {R.~L.}\ \bibnamefont {Weaver}},
  \ and\ \bibinfo {author} {\bibfnamefont {K.~A.}\ \bibnamefont {Dahmen}},\
  }\href@noop {} {\bibfield  {journal} {\bibinfo  {journal} {Phys. Rev. Lett.}\
  }\textbf {\bibinfo {volume} {117}},\ \bibinfo {pages} {261101} (\bibinfo
  {year} {2016})}\BibitemShut {NoStop}%
\bibitem [{\citenamefont {Papanikolaou}\ \emph {et~al.}(2011)\citenamefont
  {Papanikolaou}, \citenamefont {Bohn}, \citenamefont {Sommer}, \citenamefont
  {Durin}, \citenamefont {Zapperi},\ and\ \citenamefont
  {Sethna}}]{PapanikolaouETAL:2011}%
  \BibitemOpen
  \bibfield  {author} {\bibinfo {author} {\bibfnamefont {S.}~\bibnamefont
  {Papanikolaou}}, \bibinfo {author} {\bibfnamefont {F.}~\bibnamefont {Bohn}},
  \bibinfo {author} {\bibfnamefont {R.~L.}\ \bibnamefont {Sommer}}, \bibinfo
  {author} {\bibfnamefont {G.}~\bibnamefont {Durin}}, \bibinfo {author}
  {\bibfnamefont {S.}~\bibnamefont {Zapperi}}, \ and\ \bibinfo {author}
  {\bibfnamefont {J.~P.}\ \bibnamefont {Sethna}},\ }\href@noop {} {\bibfield
  {journal} {\bibinfo  {journal} {Nat. Phys.}\ }\textbf {\bibinfo {volume}
  {7}},\ \bibinfo {pages} {316} (\bibinfo {year} {2011})}\BibitemShut {NoStop}%
\bibitem [{\citenamefont {Baldassarri}\ \emph {et~al.}(2003)\citenamefont
  {Baldassarri}, \citenamefont {Colaiori},\ and\ \citenamefont
  {Castellano}}]{BaldassarriETAL:2003}%
  \BibitemOpen
  \bibfield  {author} {\bibinfo {author} {\bibfnamefont {A.}~\bibnamefont
  {Baldassarri}}, \bibinfo {author} {\bibfnamefont {F.}~\bibnamefont
  {Colaiori}}, \ and\ \bibinfo {author} {\bibfnamefont {C.}~\bibnamefont
  {Castellano}},\ }\href@noop {} {\bibfield  {journal} {\bibinfo  {journal}
  {Phys. Rev. Lett.}\ }\textbf {\bibinfo {volume} {90}},\ \bibinfo {pages}
  {060601} (\bibinfo {year} {2003})}\BibitemShut {NoStop}%
\bibitem [{\citenamefont {Willis}\ and\ \citenamefont
  {Pruessner}(2018)}]{WillisPruessner:2018}%
  \BibitemOpen
  \bibfield  {author} {\bibinfo {author} {\bibfnamefont {G.}~\bibnamefont
  {Willis}}\ and\ \bibinfo {author} {\bibfnamefont {G.}~\bibnamefont
  {Pruessner}},\ }\href@noop {} {\bibfield  {journal} {\bibinfo  {journal}
  {Int. J. Mod. Phys.}\ }\textbf {\bibinfo {volume} {32}},\ \bibinfo {pages}
  {1830002} (\bibinfo {year} {2018})}\BibitemShut {NoStop}%
\bibitem [{\citenamefont {Laurson}\ \emph {et~al.}(2013)\citenamefont
  {Laurson}, \citenamefont {Illa}, \citenamefont {Santucci}, \citenamefont
  {Tallakstad}, \citenamefont {M{\aa}l{\o}y},\ and\ \citenamefont
  {Alava}}]{LaursonETAL:2013}%
  \BibitemOpen
  \bibfield  {author} {\bibinfo {author} {\bibfnamefont {L.}~\bibnamefont
  {Laurson}}, \bibinfo {author} {\bibfnamefont {X.}~\bibnamefont {Illa}},
  \bibinfo {author} {\bibfnamefont {S.}~\bibnamefont {Santucci}}, \bibinfo
  {author} {\bibfnamefont {K.~T.}\ \bibnamefont {Tallakstad}}, \bibinfo
  {author} {\bibfnamefont {K.~J.}\ \bibnamefont {M{\aa}l{\o}y}}, \ and\
  \bibinfo {author} {\bibfnamefont {M.~J.}\ \bibnamefont {Alava}},\ }\href@noop
  {} {\bibfield  {journal} {\bibinfo  {journal} {Nat. Com.}\ }\textbf {\bibinfo
  {volume} {4}},\ \bibinfo {pages} {2927} (\bibinfo {year} {2013})}\BibitemShut
  {NoStop}%
\bibitem [{\citenamefont {Corral}\ \emph {et~al.}(2018)\citenamefont {Corral},
  \citenamefont {Garcia-Millan}, \citenamefont {Moloney},\ and\ \citenamefont
  {Font-Clos}}]{CorralETAL:2018}%
  \BibitemOpen
  \bibfield  {author} {\bibinfo {author} {\bibfnamefont {A.}~\bibnamefont
  {Corral}}, \bibinfo {author} {\bibfnamefont {R.}~\bibnamefont
  {Garcia-Millan}}, \bibinfo {author} {\bibfnamefont {N.~R.}\ \bibnamefont
  {Moloney}}, \ and\ \bibinfo {author} {\bibfnamefont {F.}~\bibnamefont
  {Font-Clos}},\ }\href {\doibase 10.1103/PhysRevE.97.062156} {\bibfield
  {journal} {\bibinfo  {journal} {Phys. Rev. E}\ }\textbf {\bibinfo {volume}
  {97}},\ \bibinfo {pages} {062156} (\bibinfo {year} {2018})}\BibitemShut
  {NoStop}%
\bibitem [{\citenamefont {{Wolfram Research
  Inc.}}(2018)}]{Mathematica:11.3.0.0}%
  \BibitemOpen
  \bibfield  {author} {\bibinfo {author} {\bibnamefont {{Wolfram Research
  Inc.}}},\ }\href@noop {} {\emph {\bibinfo {title} {Mathematica}}}\ (\bibinfo
  {publisher} {Wolfram Research, Inc.},\ \bibinfo {address} {Champaign, IL},\ 
  \bibinfo {year} {2018}),\ \bibinfo {note} {version
  11.3.0.0}\BibitemShut {NoStop}%
\bibitem [{\citenamefont {Callan}(2009)}]{callan:2009}%
  \BibitemOpen
  \bibfield  {author} {\bibinfo {author} {\bibfnamefont {D.}~\bibnamefont
  {Callan}},\ }\href@noop {} {\bibfield  {journal} {\bibinfo  {journal} {arXiv
  preprint arXiv:0906.1317}\ } (\bibinfo {year} {2009})}\BibitemShut {NoStop}%
\end{thebibliography}%

\end{document}